\documentclass[journal,9pt]{IEEEtran}
\ifCLASSINFOpdf
\else
\fi

\usepackage[cmex10]{amsmath}
\usepackage{graphicx}
\usepackage{caption}
\usepackage{subfigure}
\usepackage{multicol}
\usepackage{multirow}
\usepackage{soul}

\usepackage{lipsum}
\usepackage{adjustbox}
\usepackage[figuresright]{rotating}
\usepackage{algorithmicx}
\usepackage{algorithm2e}
\usepackage{amsmath}

\usepackage[usenames, dvipsnames]{color}

\usepackage[table]{xcolor}
\setulcolor{blue}

\setstcolor{red}

\soulregister\cite7
\soulregister\ref7
\soulregister\pageref7

\makeatletter
\AtBeginDocument{\let\hl\@firstofone}
\makeatother

\begin{document}
%
\title{Modeling Impact of Human Errors on the Data Unavailability and Data Loss of Storage Systems}
%
%
%

\author{\IEEEauthorblockN{{ Mostafa Kishani 
and Hossein Asadi, Senior Member, IEEE\\}}
\IEEEauthorblockA{{ Data Storage, Networks, \& Processing (DSN) Lab, Department of Computer Engineering, \\Sharif University of Technology}}}

\maketitle

\begin{abstract}
Data storage systems and their availability play a crucial role in contemporary datacenters. Despite using mechanisms such as automatic fail-over in datacenters, the role of human agents and consequently their destructive errors is inevitable. Due to very large number of disk drives used in exascale 
datacenters and their high failure rates, the disk subsystem in storage systems has become a major source of \emph{Data Unavailability} (DU) and \emph{Data Loss} (DL) initiated by human errors.
In this paper, we investigate the effect of \emph{Incorrect Disk Replacement Service} (IDRS) on the availability and reliability of data storage systems.
To this end, we analyze the consequences of IDRS in a disk array, 
and conduct Monte Carlo simulations to evaluate DU and DL during mission time.  
The proposed modeling framework can cope with a) different storage array configurations and b) \emph{Data Object Survivability} (DOS), representing the effect of system level redundancies such as remote backups and mirrors. 
In the proposed framework, the model parameters are obtained from industrial and scientific reports alongside field data which have been extracted from a datacenter operating with 70 storage racks.
The results show that ignoring the impact of IDRS leads to unavailability underestimation by up to three orders of magnitude. 
Moreover, our study suggests that by considering the effect of human errors, the conventional beliefs about the dependability 
of different \emph{Redundant Array of Independent Disks} (RAID) mechanisms should be revised. 
The results show that $RAID1$ can result in 
lower availability compared to $RAID5$ in the presence of human errors.
The results also show that employing automatic fail-over policy (using hot spare disks) can reduce the drastic impacts of human errors by two orders of magnitude. 

\end{abstract}


%
\IEEEpeerreviewmaketitle

\section{Introduction}
\label{sec:Intro}
The availability and reliability of Information systems is seriously affected by human errors
~\cite{oppenheimer2003internet,brown2001err,oppenheimer2003importance,kishani2017} 
where some field studies report human errors as the cause of 19\% of system failures~\cite{haubert,oppenheimer2003importance}. 
Large datacenters with \emph{Exa-Byte} (EB) storage capacity (by employing millions of disks drives) are expected to face at least 
a disk failure per hour.
Mechanisms such as automatic fail-over try to reduce the role of human agent in service and maintenance tasks, however, 
in many cases the involvement of human is inevitable. 
Meanwhile, 
despite precautionary mechanisms such as using checklists and complying high standards for training the technicians, 
the \emph{human error probability} ($hep$) is between 0.001 and 0.1
~\cite{NASA-her,gibson2006feasibility,us1975reactor,swain1983handbook}. 
These statistics translate into multiple human errors a day in an exascale datacenter. 
As a simple and frequent example of human error in data-centers, assume 
an array with one failed disk, and a human agent that is responsible for replacing the failed disk with the brand-new one.
However, due to the lack of concentration, he or she wrongly removes the operating disk, rather than the failed one.
This makes the whole array unavailable and can even lead to data loss if the wrongly replaced disk is thrown away~\cite{kishani2017}.  

The most vulnerable component in a \emph{Data Storage System} (DSS)\footnote{Data storage system is responsible to retain digital data with a higher reliability and performance level than individual storage medias, by mechanisms such as caching~\cite{tarihi2016hybrid,reca,ahmadian2018eci}, data tiering~\cite{salkhordeh2015operating}, and redundancy.} is disk drive, where 
disk failures and \emph{Latent Sector Errors}\footnote{Damages to disk sectors, caused by bad head writes, bit errors, and environmental particles which may be placed between platter and head.} (LSE)~\cite{Schroeder-2010-TOS} cause the majority of 
\emph{Data Loss} (DL) in data-centers.
Investigating the effect of these two incidences on the reliability of disks drives and disk arrays have been the  subject of several studies 
~\cite{elerath2014beyond,elerath-DSN-2007,Schroeder-2010-TOS,Elerath-2009-TC,Greenan-HOTSTORAGE-2010,Schroeder-FAST-2007,Elerath-DSN-2009,elerath2016raid,ma2015raidshield,paris2014protecting}. 
Elerath and Pecht~\cite{elerath-DSN-2007,Elerath-2009-TC} show that the conventional reliability estimation approach, \emph{Mean Time to Data Loss} (MTTDL), can result in DL underestimation by orders of magnitude, as using MTTDL approach mandates assuming exponential distribution for both disk failure and fail-over rates, which is not realistic. 
In return, this study leverages the field data and shows that the rate of operational disk failure, LSE, disk fail-over, and 
\emph{Disk Scrubbing}\footnote{A task that removes LSEs by periodically reading the disk data and checking it with its parity, correcting the corrupted data using the parity and moving it to a new location, and mapping out the damaged sectors.} can follow a three-parameter Weibull distribution. 
This work evaluates the reliability of \emph{Redundant Array of Independent Disks} (RAID) using Monte Carlo simulations, but arguably takes the loss of one data stripe (by LSE) as a \emph{Double Disk Failure}\footnote{An event in which the whole data of $RAID5$ array is lost, due to the consecutive failure of two disks.} (DDF) and finally counts the number of DDFs as a reliability metric, which results in data loss overestimation. 
Moreover, Elerath and Pecht~\cite{elerath-DSN-2007,Elerath-2009-TC} just consider the single configuration of array having infinite cold-spares (mandating human assistance  in disk fail-over), while  ignoring the effect of human errors.
Greenan et. al.~\cite{Greenan-HOTSTORAGE-2010} proposes \emph{NOrmalized Magnitude of Data Loss} (NOMDL) metric, defined as the amount of data loss within mission time, normalized to the usable capacity of disk array, to cope with the limitations of DDF metric.  
Elerath and Schindler~\cite{elerath2014beyond} extend the $RAID5$ models appeared in~\cite{elerath-DSN-2007,Elerath-2009-TC,Elerath-DSN-2009} to be applied 
to $RAID6$ arrays, by proposing a closed-form equation that uses a table of failure and repair parameters obtained by Monte Carlo simulations using Weibull distribution. 
One can conclude that the focus of all previous work is on DL in the disk array, ignoring the possibility of \emph{Data Unavailability} (DU) caused by human errors.

Considering the effect of human errors alongside the knowledge provided by previous models and field studies, we can conclude that an accurate modeling of 
RAID dependability is very crucial to take into account several important criteria including 
a) a realistic distribution for failure and repair rates, 
b) the effect of LSEs and its differences with operational disk failures, 
c) the possibility of human errors in array service and maintenance, 
and d) evaluation of both reliability and availability within mission time while considering fair and meaningful metrics for reliability and availability.
To the best of our knowledge, none of previous studies have addressed these concerns in a unified framework, while the effect of human errors is totally missed in the previous dependability models.

In this paper, we propose a dependability model for the disk arrays by considering 
the effect of disk failures, LSEs, and \emph{Incorrect Disk Replacement Service} (IDRS) as a common sample of human errors\footnote{While the incorrect repair service can have many different roots and happen in many different conditions, in this work we focus on IDRS.}.
To this end, we analyze the possible combinations of operational disk failures, LSEs, and IDRS in a disk array. 
This analysis which is demonstrated by state diagrams, concludes that the combination of disk failure and IDRS can result in the unavailability of the whole array, while the combination of LSE and IRDS results in the unavailability of one or multiple data stripes, mandating a metric which is capable to project the magnitude of data unavailability as well as unavailability duration.
We further define \emph{NOrmalized Magnitude of Data Unavailability} (NOMDU), as the duration of data unavailability multiplied to the amount of unavailable data (in an arbitrary unit such as mega bytes) within mission time, normalized to the mission time and usable capacity of disk array. 
In our analysis, both disk subsystems with and without automatic disk fail-over are considered. 
 
Using the proposed failure analysis, we conduct Monte Carlo simulations to evaluate NOMDU and NOMDL during mission time, by considering three-parameter 
Weibull distributions for the rate of operational disk failure, LSE, and IDRS, as well as the corresponding repair rates.   
Several important observations are obtained by the proposed model.
First, 
it is shown that human errors can result in storage unavailability by order of magnitude (up to $NOMDU=10^{-5}$ when human error probability is 0.1). 
The human error can also increase the probability of data loss, specially when the human error probability is more than 0.01 (human error probability of 0.1 can increase data loss by one order of magnitude). 
Second, the presence of human errors can contradict the conventional assumption about the dependability of RAID mechanisms, 
as the RAID configurations with greater level of redundancy suffer higher unavailability caused by human errors.  
Third, it is demonstrated that automatic disk fail-over, when on-line rebuilt is provided by using spare disks, can reduce the drastic impacts of human errors by orders of magnitude. 

The model parameters are obtained from industrial and scientific reports alongside field data, which are extracted  from the main datacenter of \emph{Sharif University of Technology} (SUT)\footnote{This data-center offers various Cloud-based services, web-hosting, collocation, mail service, and HPC services to both universities and small to medium-size corps.}~\cite{sharifuni}, operating with 70 storage and computing racks (with more than 100PB storage capacity).
This datacenter is equipped with \emph{SAB-SE}~\cite{SAB-SE} storage nodes\footnote{A modular DSS designed and fabricated by HPDS Corp.~\cite{HPDScorp}.} each of which supporting up to 72 disk drives, enabling the datacenter to support more than 27,000 disk drives.

\textbf{Our contribution over the recent work ~\cite{kishani2017} is as follows:}
\begin{itemize}
\item
The proposed model is extended to consider the effect of a) LSEs for $RAID5$ arrays and b) $RAID5$ with spare disk.
\item
Models in ~\cite{kishani2017} assume a 100\% survivable storage system\footnote{A data object, stored in a DSS, is called survivable if it has a backup or remote mirror, enabling data recovery in the case of local data loss~\cite{li2012understanding}. Otherwise, it is called non-survivable.}, while this work assumes the general case in which parts of data can be non-survivable.
\item
For the first time, a novel metric, NOMDU, is proposed to assess the availability of data storage systems.
\item
By considering the data object survivability as a model parameter, the proposed model reports availability and reliability in terms of NOMDU and NOMDL.
\item
Monte Carlo simulation is used to assess NOMDU and NOMDL, rather than Markov models, while time-to-failure and time-to-repair is generated by considering Weibull 
distribution, obtained from field data and state-of-the-art reports. 
\item
Model presentation is revised to improve its understandability and applicability.  
\end{itemize}

The remainder of this paper is organized as follows.
Section~\ref{sec:related} represents background and related works. 
Section~\ref{sec:ref-avail} elaborates the human error analysis in disk arrays using Monte-Carlo simulations. 
Section~\ref{sec:results} provides simulation results 
and the corresponding findings. 
Lastly, Section \ref{sec:Conclude} concludes the paper.

\section{Background and Related Work}
\label{sec:related}

\subsection{Dependability Models of Data Storage Systems}

Many research studies have tried to evaluate and improve the reliability of data storage systems (in particular, disk subsystem) by considering the
failure cases that result in data loss
~\cite{Gibson-BOOK-1990,Blaum-TC-95,Elerath-DSN-2009,Elerath-2009-TC,Greenan-HOTSTORAGE-2010,elerath2016raid,ma2015raidshield,paris2014protecting,venkatesan2012general,ahmadian-ssd-rel-date}. 
Metrics of data reliability used in the literature include a) MTTDL~\cite{Gibson-BOOK-1990} which attempts to express the average time between data loss events, 
b) DDF~\cite{Blaum-TC-95,Elerath-DSN-2009,Elerath-2009-TC,elerath2016raid} which expresses the expected time between failures, 
c) percentage of RAID array failures within mission time~\cite{ma2015raidshield}, 
and d) \emph{Magnitude of Data Loss} (MDL)~\cite{Greenan-HOTSTORAGE-2010} which is the amount of data (in bytes) that is expected to be lost within mission time.
The other dependability parameter, data availability, expresses the fraction of time that data 
is accessible by customers~\cite{avizienis-DSC-2004}. 
Dependability of data storage systems can be significantly influenced by parameters such as the rate of component failures, 
the rate of recovery mechanisms, and the structure of redundancy mechanism used to tolerate component failures. 
A variety of redundancy and recovery techniques is employed in data storage systems 
to mitigate the consequences of component failures and decrease the probability of data unavailability and/or data loss.
These mechanisms usually come with considerable performance, energy consumption, or cost overheads.
Hence, designers manage to use system-level dependability models to measure the effectiveness of redundancy mechanisms applied to data storage systems reaching cost-effective redundancy techniques.

\subsection{Human Error in Safety-Critical Applications}
\label{sec:human_error}

\emph{Human Reliability Assessment} (HRA) ~\cite{swain1990human} techniques are developed 
to attain a better understandability and quantification of human errors in a non-benign system. 
These techniques mainly focus on quantifying $hep$ which is simply defined by Equation \ref{equ:hep} ~\cite{gibson2006feasibility}.

\begin{equation}
\label{equ:hep}
\begin{centering}
hep=\frac{No.~of~error~cases~observed}{No.~of~opportunities~for~human~errors}
\end{centering}
\end{equation}

By referring to $hep$ values obtained by \emph{National Aeronautics and Space Administration} (NASA), \emph{European Organization for the Safety of Air Navigation} (EUROCONTROL), and \emph{United States Nuclear Regulatory Commission} (NUREG), it can be concluded that the probability of human error is usually between $0.001$ and $0.1$ depending on the application and situation.
However, for the most of safety-critical and enterprise applications, the reported $hep$ is in the range of $0.001$ and $0.01$ 
\cite{gibson2006feasibility,us1975reactor,swain1983handbook,NASA-her}.

Finally, we can note studies inspecting and modeling the effect of human errors in enterprise systems such as nuclear power plants
~\cite{dhillon1983system,apostolakis1977effect}, 
and studies trying to improve the maintenance and test quality of enterprise systems, in favor of maintenance cost and reliability/availability
~\cite{berrade2015some,mcwilliams1980human}.

\subsection{Human Errors in Data Storage Systems}
\begin{figure}
\begin{centering}
\includegraphics[width=2.2in]{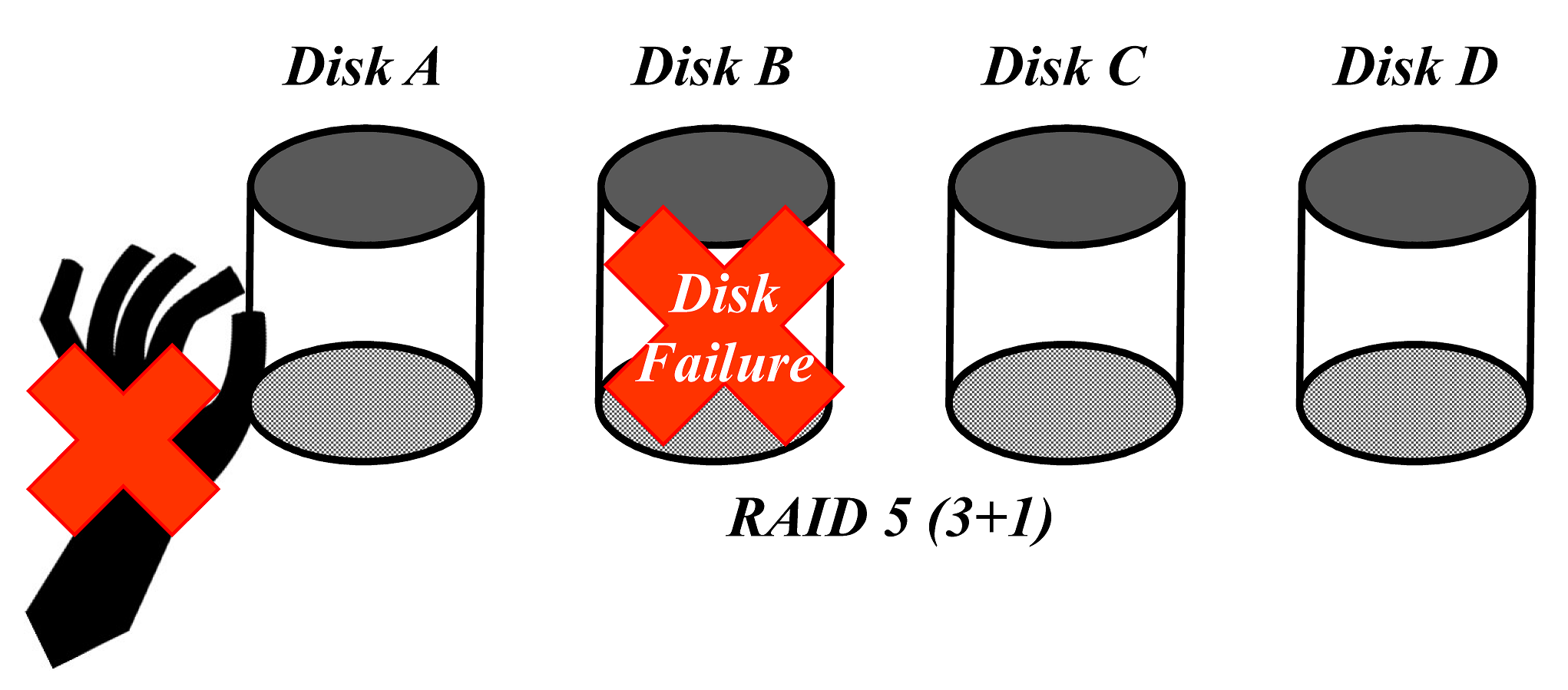}
\caption{An example: a human error in disk replacement process can result in data unavailability in the disk array.}
\label{fig:human error sample}
\par\end{centering}
\vspace{-0.3cm}
\end{figure}

Human errors can threat the availability/reliability of DSSs in different components and situations, however, 
in this work we investigate the effect of IDRS which is one of the most prevalent types of human errors.
Consider a $RAID5$ array with no spare disks, in which the failed disk should be replaced by the brand-new disk 
before starting the fail-over process. 
As shown in Fig.~\ref{fig:human error sample}, 
the operator may wrongly replace the brand-new disk with one of the operating disks, rather than the failed one.
This incidence, called IDRS, makes two disks, the wrongly removed one and the failed one, inaccessible, resulting in 
the unavailability of the entire array. 
If the human error is detected, the array will be available by undoing the incorrect disk replacement. 
Otherwise, if the wrongly removed disk is damaged before the detection and recovery of human error, 
the entire array will be lost due to DDF.

\subsection{LSE}
\label{sec:lse}
Most studies in the field of data storage systems have focused on the failure analysis of disks, including operational failures and 
undetected errors~\cite{elerath-DSN-2007,Rozier-DSN-2009,Bairavasundaram-ACM-2007,pinheiro2007failure}. 
Operational failures occur due to faults in electronic and mechanical components such as heads and platters.
These failures result in data destruction where disk head is unable to find the requested data. 
Using RAID configurations is a solution for alleviating the effect of operational failures on data storage systems~\cite{Patterson-ACM_SIGMOD-1988}. 

\begin{figure}
\begin{centering}
\includegraphics[width=2in]{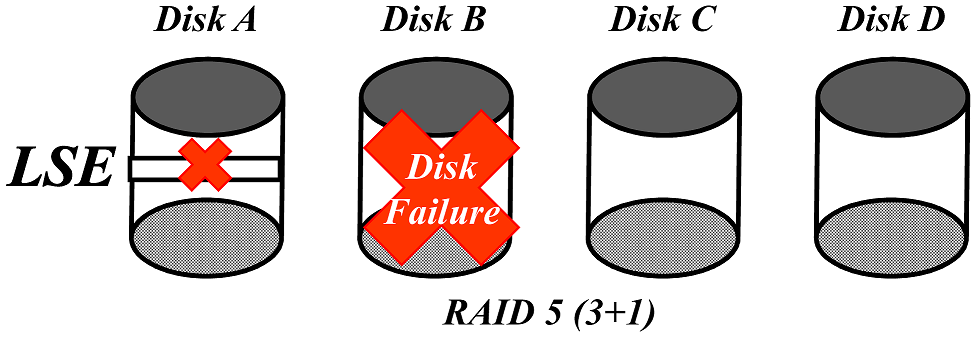}
\caption{An example: an LSE followed by a disk failure can result in data loss in the LSE-affected sectors.}
\vspace{-0.3cm}
\label{fig:lse-sample}
\par\end{centering}
\vspace{-0.3cm}
\end{figure}

In addition to operational failures, other types of errors such as bad head write and bit error can also damage disk sectors. 
Another cause of sector errors is environmental particles which may be placed between platter and head. 
In the case of a write operation, positioning the disk head within track gaps can corrupt several sectors. 
These types of errors, named LSEs, 
may lead to a data loss event upon a disk failure~\cite{Bairavasundaram-ACM-2007,Schroeder-2010-TOS}. 
Fig.~\ref{fig:lse-sample} shows how an LSE can result in data loss in the case of a subsequent disk failure in the case of $RAID5$.
Suppose that a sector of disk $A$ is affected by LSE. If disk $B$ fails before detection of recovery of LSE, the data of affected sector in disk $A$ cannot 
be recovered, as $RAID5$ can just tolerate the failure of one disk.
\emph{Error Correcting Code} (ECC)~\cite{Li-2009-GRID,Li-2011-SIGMETRICS}, disk scrubbing~\cite{Mi-DSN-2008}, and intra-disk redundancy \cite{dholakia-TOS-2008,Iliadis-ACM_SIGMETRICS-2008} can be used 
to reduce the probability of data corruption in the presence of LSEs.
The LSE rate of a disk drive may vary in time, depending on several parameters such as disk age, disk model, and I/O characteristics~\cite{Bairavasundaram-ACM-2007}.
\hl{We should note that some works on disk array reliability (such as Venkatesan and Iliadis~\cite{venkatesan2012general}) ignore the effect of LSEs that results in misleading conclusions~\cite{Greenan-HOTSTORAGE-2010,greenan2009reliability,elerath-DSN-2007,Schroeder-2010-TOS,Schroeder-FAST-2007}.}

\section{Human Error Analysis in Disk Array}
\label{sec:ref-avail}
\hl{In this section, we model the dependability of disk arrays using Monte Carlo simulations rather than conventional alternatives such as MTTDL and Markov models due to their extensive limitations and inaccuracies.  
Many previous studies have concluded that MTTDL is an obsolete metric for reporting Data Loss~\cite{brown2001err,Greenan-HOTSTORAGE-2010,rao2006reliability,rosenthal2010bit}.
The disk arrays have infinite failure states (due to having infinite combinations of sector failures, disk failures, and human errors) and modeling them with a closed-form MTTDL expression, and even a Markov chain is challenging and erroneous~\cite{e2000transient}. 
Furthermore, the disk failure rate is a function of time that makes using Markov chains erroneous~\cite{Greenan-HOTSTORAGE-2010,greenan2009reliability}, while many previous works encourage using alternatives such as Monte Carlo simulations that have not this limitation~\cite{brown2001err,Greenan-HOTSTORAGE-2010,greenan2009reliability,elerath-DSN-2007,Schroeder-2010-TOS,Schroeder-FAST-2007,Bairavasundaram-ACM-2007}. 
In this section, we first introduce NOMDU metric for evaluating the availability of data storage systems. Afterwards, we propose our framework for evaluating the dependability of $RAID5$ and $RAID6$ arrays by considering disk failures, LSEs, and human errors. 
Finally, we discuss the dependability of general erasure codes and how our proposed framework can be employed for different code configurations. 
}

\subsection{NOrmalized Magnitude of Data Unavailability (NOMDU)}
\label{sec:nomdu}
To access unavailability in a data storage systems, we need a metric to be \hl{applicable and} comparable in different storage capacities, and \hl{contain the magnitude of unavailable data}. 
The original availability/unavailability metric cannot be useful in the case of storage systems, for two reasons:

\textbf{Case A)} Availability is a function of storage capacity, \hl{while a storage system with a larger capacity but the same architecture will have lower availability. Hence, different storage architectures with different capacities cannot be compared using DU metric. 
We take an example where two system engineers evaluate the availability of two storage subsystems using conventional availability metric. Assume \emph{Subsystem 1} ($SS1$) employs one $RAID0(4~disks)$ array and \emph{Subsystem 2} ($SS2$) employs two $RAID0(4~disks)$ arrays, while the arrays of both subsystems have exactly the same architecture and components.  
Assume $A_{Disk}$ stands for the availability of each disk, $A_{array}$ stands for the availability of one disk array, and $A_{SS1}$ and $A_{SS2}$ respectively stand for the availability of $SS1$ and $SS2$.
Regarding $RAID0$ configuration, the array is unavailable when at least one of disks is unavailable.
Moreover, in the conventional availability definition, when one of two arrays is unavailable, the whole system is considered unavailable (as unavailability metric does not deliver any information about the magnitude of data unavailability).
In summary, conventional availability of $SS1$ and $SS2$ is as follows:
$A_{SS1} = A^{4}$,~~~~~~~~$A_{SS2}=A^{8}$
}

\hl{As the formulations of $A_{SS1}$ and $A_{SS2}$ show, the conventional availability is a function of system scale. Hence, two systems with exactly the same architecture but different scales have different availability values. Moreover, the availability does not change linearly with system capacity (system scale). Hence, the system engineers cannot obtain the availability of $SS1$, by simply normalizing the availability of $SS2$ to its capacity. }

\textbf{Case B)} Unavailability \hl{metric} cannot represent the magnitude of unavailable data.
In many failure cases, only a part of storage data is unavailable, while the definition of storage availability/unavailability is limited to the availability of whole data (the storage is considered available when the whole data is available).
\hl{We take an example to elaborate this shortcoming of availability metric when used in data storage systems. To this end, we evaluate the conventional availability of a data storage system employing a single HDD and a true remote backup (such as Cloud backup). 
Suppose two failure types of disk failure and LSE are possible in a HDD with the following definitions: a) disk failure: \emph{Time To Failure} (TTF), \emph{Time To Recover} (TTR), and b) LSE: \emph{Time Between LSE} (TBLSE), \emph{Time To LSE Recover} (TTLSER). 
The storage system is available when all its data is available, i.e., when no unavailability is caused by disk failure and LSE:}
\resizebox{1\hsize}{!}{$A_{DSS} = A_{DSS}(Disk Failure) \times A_{DSS}(LSE) = \frac{TTF}{TTF+TTR} \times \frac{TBLSE-TTLSER}{TBLSE}$}

\hl{The shortcoming of conventional availability metric, as shown in above formulation, is that both HDD failure and LSE have the same impact on system availability, while they cause totally different magnitude of data unavailability (the whole disk size versus a single sector size).}
 
Here we define NOMDU, 
as the duration of data unavailability multiplied to the logical amount of unavailable data, normalized to the mission time and logical capacity of storage system, as shown in Equation~\ref{equ:nomdu-original}. 
Hence, this metric can assess the availability of a storage architecture, regardless of its size and mission time. 
\vspace{-0.3cm}

\begin{equation}
\label{equ:nomdu-original}
\begin{centering}
\resizebox{1\hsize}{!}{$NOMDU=\frac{\sum Logical~Size~of~Unavailable~Data \times Unavailability~Duration}{Total~Logical~Storage~Size \times Mission~Time}$}
\end{centering}
\end{equation}

\hl{Following we calculate NOMDU for \textbf{Case A} and \textbf{Case B} (appeared above) to demonstrate how NOMDU removes the problems of conventional availability metric. }

\hl{\textbf{Case A)} 
$NOMDU_{SS1}=NOMDU_{SS2}=1-A$ }

\hl{Regarding \textbf{Case A}, two systems with the same architecture but different scale have the same NOMDU, while they have different conventional availability.}

\hl{\textbf{Case B)}
$NOMDU = \frac{Capacity_{sector}}{Capacity_{disk}} \times \frac{TTLSER}{TBLSE} + \frac{TTR}{TTF+TTR}$ }

\hl{As the NOMDU formulation for \textbf{Case B} shows, the unavailability caused by LSE and disk failure have different impact on NOMDU, while their impact is proportional to the fraction of their capacity over total storage capacity. }

\subsection{Dependability of $RAID5$ and $RAID6$, No LSE, No Automatic Fail-over}
\label{sec:analysis-no lse-no spare}

\subsubsection{RAID5 Analysis}
\label{sec:raid5-no lse-no spare}

\begin{figure}
\begin{centering}
\includegraphics[width=3in]{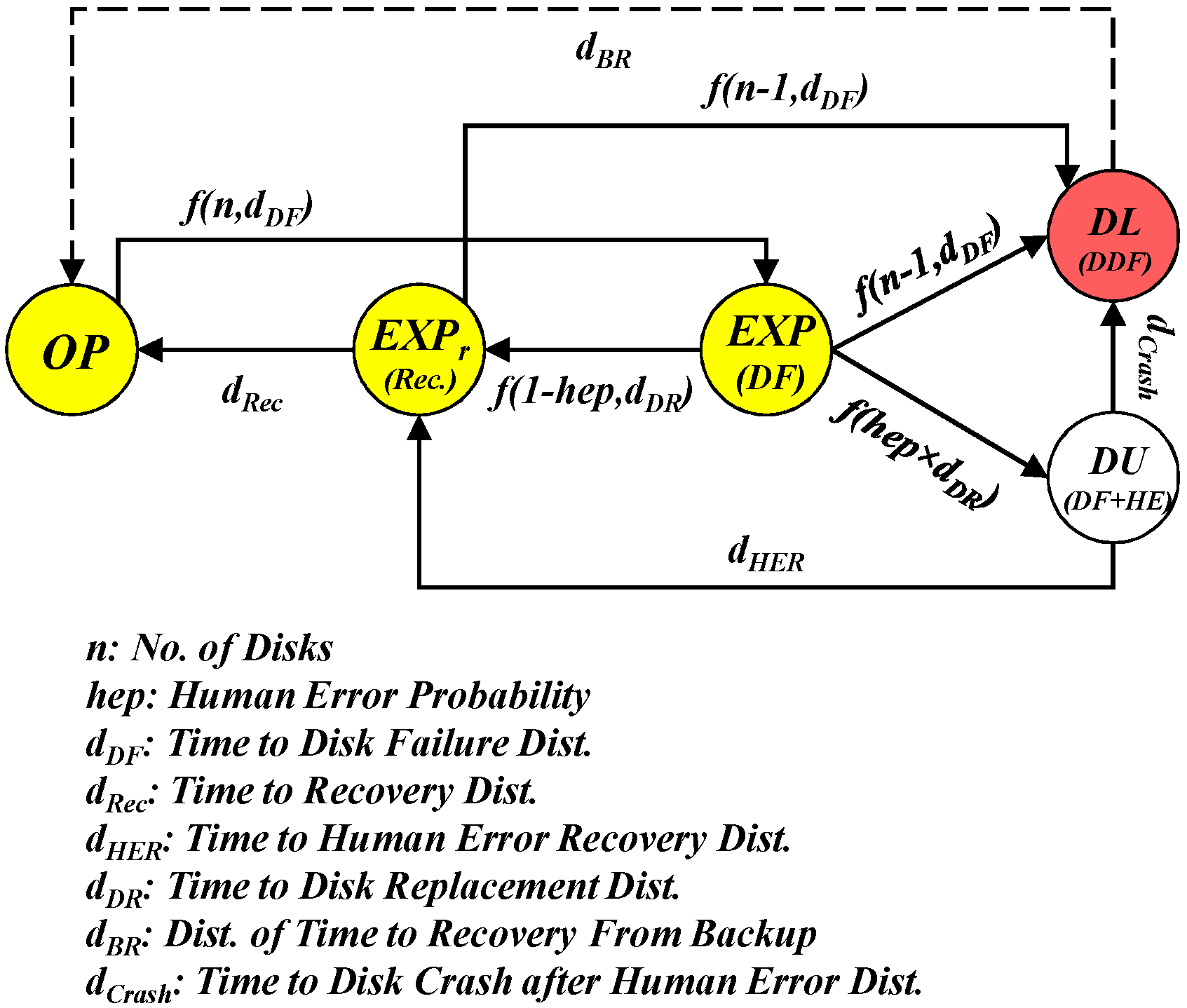}
\caption{State diagram of Monte Carlo simulation for $RAID5$ DU/DL, considering no LSE.}
\label{fig:raid5hr}
\par\end{centering}
\vspace{-0.5cm}
\end{figure}

Fig.~\ref{fig:raid5hr} shows the proposed state diagram for assessing DU/DL in a $RAID5$ disk subsystem by 
considering the effect of disk failures and human errors. 
This model is evaluated using Monte Carlo simulations, as using Markov models can be erroneous due to its memoryless nature that prevents modeling 
non-exponential failure distributions such as Weibull~\cite{thompson1988rate,elerath-DSN-2007}.

We have the same convention in naming the states in all state diagrams. The states in which the next failure results in DU/DL are named EXP and the states in which the next failure does not result in DU/DL are named OP.  
Upon the occurrence of the first disk failure, 
the system state will move from the operational ($OP$) to the exposed state ($EXP$).
While being in the exposed state, 
a second disk failure will lead to DL event 
whereas a human error during disk replacement will lead to DU event.
If the human agent successfully replaces the failed disk with the brand-new one, the array goes to the $EXP_r$ state, in which the disk fail-over can be started on the brand-new disk.

When the array is in the $DU$ state, by recognizing the human error and removing it, the array switches to the $EXP_r$ state, in which the failed disk is correctly replaced by the brand-new one and the fail-over process can be started. 
However, if the wrongly replaced disk is crashed, a DDF happens and the array switches to the $DL$ state. 
The time to crash the wrongly replaced disk is considered to have the distribution of $d_{crash}$.
Per DU incidence $i$, NOMDU is evaluated using Equation~\ref{equ:nomdu} and is added to the simulation statistics.
\vspace{-0.3cm}

\begin{equation}
\label{equ:nomdu}
\begin{centering}
\resizebox{1\hsize}{!}{$NOMDU_{i}=\frac{Logical~Size~of~Unavailable~Data_{i} \times Unavailability~Duration_{i}}{Total~Logical~Storage~Size \times Mission~Time}$}
\end{centering}
\end{equation}

In this regard, Equation~\ref{equ:nomdu-original} is rephrased as follows:

\begin{equation}
\label{equ:nomdu-original-2}
\begin{centering}
NOMDU=\sum_{i} NOMDU_{i} 
\end{centering}
\end{equation}

Where $NOMDU$ is normalized magnitude of data unavailability within mission time, and $NOMDU_{i}$ is NOMDU imposed by DU incidence $i$. 

Finally, when the array is in the $DL$ state, the whole array data is lost due to DDF. 
In a non-survivable storage, that has no backup and mirror, in this case the array data is permanently lost.
Hence, NOMDL is evaluated and added to the simulation statistics as shown in Equation~\ref{equ:nomdl}:

\begin{equation}
\label{equ:nomdl}
\begin{centering}
NOMDL_{nonsurvivable_i}=\frac{Logical~Size~of~Lost~Data_{i}}{Total~Logical~Storage~Size}
\end{centering}
\end{equation}

Where $NOMDL_{nonsurvivable_i}$ is normalized magnitude of data loss imposed by non-survivable DL incidence $i$. 
In this regard, NOMDL within mission time is the aggregation of NOMDL imposed by individual DL incidence, as shown in Equation~\ref{equ:nomdl-original}

\begin{equation}
\label{equ:nomdl-original}
\begin{centering}
NOMDL=\sum_{i} NOMDL_{i} 
\end{centering}
\end{equation}

In the case of DL in a survivable storage, that has at least one up-to-date backup or mirror, the array data can be recovered from the backup. 
In this case it takes \emph{Backup Recovery Time}, $d_{BR}$ to recover the data of lost array over the remote backup, while \emph{Backup Recovery Time} depends on the parameters such as the size of lost data, backup throughput, array throughput, and network bandwidth. 
The survived data is not lost in the user side, but is unavailable within recovery time. 
Hence, NOMDU imposed by survivable DL incidence $i$ is evaluated as Equation~\ref{equ:nomdu-survivable}.
\vspace{-0.3cm}

\begin{equation}
\label{equ:nomdu-survivable}
\begin{centering}
\resizebox{1\hsize}{!}{$NOMDU_{survivable\_ DL_i}=\frac{Logical~Size~of~Lost~Data_{i} \times Recovery~Time_{i}}{Total~Logical~Storage~Size \times Mission~Time}$}
\end{centering}
\end{equation}

In general, we can consider \emph{Data Object Survivability} (DOS)~\cite{li2012understanding}, 
defined as the probability that a data object is survived during period of time (t).
\hl{$DOS(t)$} can be statistically interpreted as follows. 
Per DL incidence at the storage system level, a fraction of lost data, \hl{$DOS(t)$}, has a correct backup at mission time $t$, while the rest of data (\hl{$1-DOS(t)$} fraction of data) has no correct backup and is permanently lost. NOMDL metric is projecting the data that is permanently lost in the user side. Hence, in each DL incidence, NOMDL is a function of DL magnitude (size of lost data at the storage system level) and \hl{$1-DOS(t)$}, i.e., the fraction of data that has no correct backup, as shown in Equation~\ref{equ:nomdl-dos}. 
Moreover, in each DL incidence, \hl{$DOS(t)$} fraction of data is not permanently lost, as it is recoverable from remote backups and mirrors. This fraction of data is just unavailable (DU in the user side) within recovery time. Hence, imposed NOMDU per DL incidence is a function of DL magnitude (the size of lost data at the storage system level), \hl{$DOS(t)$}, and DL recovery time (from backup), as shown in Equation~\ref{equ:nomdu-dos}.

\vspace{-0.3cm}

\begin{equation}
\label{equ:nomdu-dos}
\begin{centering}
\resizebox{1\hsize}{!}{$NOMDU_{DL_i}=DOS(t) \times \frac{Logical~Size~of~Lost~Data_{i} \times Recovery~Time_{i}}{Total~Logical~Storage~Size \times Mission~Time}$}
\end{centering}
\vspace{-0.3cm}
\end{equation}

\begin{equation}
\label{equ:nomdl-dos}
\begin{centering}
NOMDL_i=(1-DOS(t)) \times \frac{Logical~Size~of~Lost~Data_{i}}{Total~Logical~Storage~Size}
\end{centering}
\end{equation}

Finally, total NOMDL and NOMDU per mission is evaluated respectively by the aggregation of NOMDL and NOMDU within mission time, as shown in Equation~\ref{equ:nomdl-original} and Equation~\ref{equ:nomdu-original-2}, respectively.

\subsubsection{RAID6 Analysis}

\begin{figure*}
\begin{centering}
\includegraphics[width=7in]{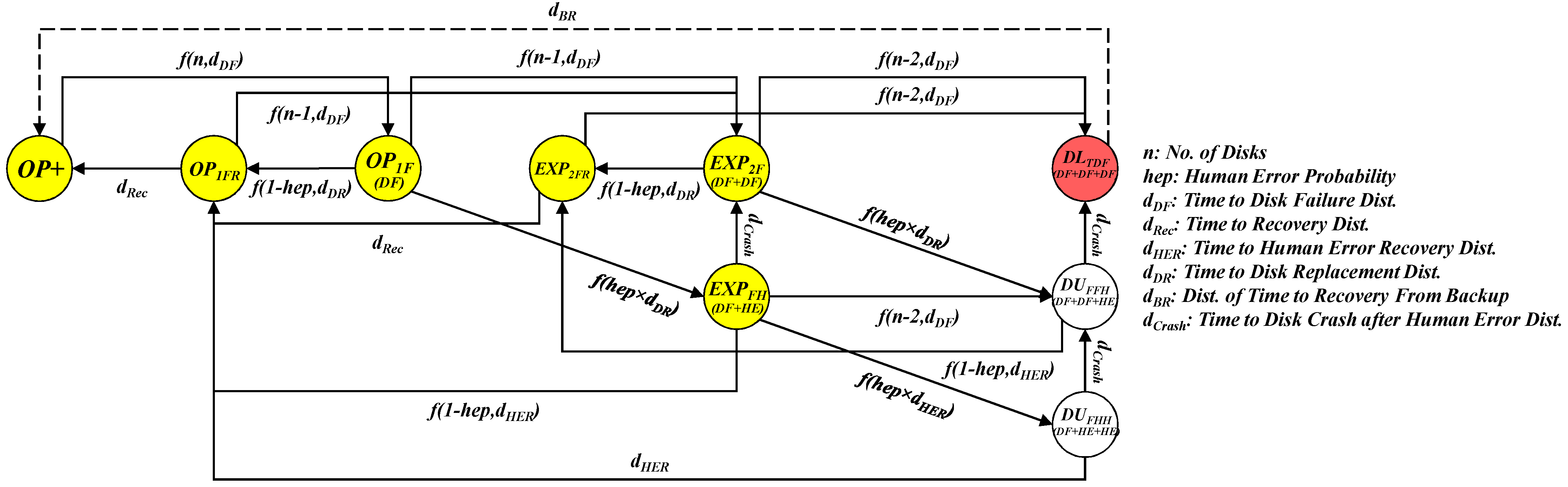}
\caption{State diagram of Monte Carlo simulation for assessing $RAID6$ DU/DL.}
\label{fig:raid6hr}
\par\end{centering}
\vspace{-0.3cm}
\end{figure*}

The proposed model for $RAID5$ (Fig.~\ref{fig:raid5hr}) is extended to assess DU/DL of a $RAID6$ array in the presence of human errors and disk failures,
as shown in Fig.~\ref{fig:raid6hr}. 
In the $RAID6$ configuration, two redundant disks are used to tolerate two consecutive disk failures. 
Hence, the data loss event happens in the case of \emph{Triple Disk Failure} (TDF).
In the normal operation of a $RAID6$ array (shown as $OP+$ state in Fig.~\ref{fig:raid6hr}), 
one and two disk failures will bring the array to either $OP_{1F}$ and $EXP_{2F}$ states, respectively. 
$OP_{1F}$ stands for the state in which one disk is failed, but the array is still operational. In this state, another disk failure moves the array to the $EXP_{2F}$ state. In the $OP_{1F}$ state, a successful disk replacement moves the array to the $OP_{1FR}$ state, while an unsuccessful disk replacement moves the array to $EXP_{FH}$ state. 

The exposed state in this figure expresses that the array will continue servicing read/write requests.
However, in the case of another disk failure before the performing the recovery process, a TDF happens that results in DL.
Assessing NOMDU and NOMDL for each DU and DL incidence is similar to the case of $RAID5$ (Section~\ref{sec:raid5-no lse-no spare}).
While the array is in the exposed state, a wrong disk replacement can make the array unavailable. 
Additionally, while the array is in $OP+$, a single disk failure followed by two consecutive wrong disk replacements can make the array unavailable.
If the disk replacement is performed with no human error, the array
goes back to either $OP_{1FR}$ and $EXP_{2FR}$ states, when it is in $OP_{1F}$ and $EXP_{2F}$ states respectively, 
while the occurrence of a human error in the disk replacement changes the array state to either $EXP_{FH}$ and $DU_{FFH}$ states. 
The array goes to $DU_{FFH}$ state when the combination of two disk failures and one human error happens and goes to $DU_{FHH}$ state when a disk failure is followed by two human errors.

$EXP_{2F}$ stands for the state in which two successive disk failures happen. In this state, another disk failure moves the array to $DL_{TDF}$ (triple disk failure). 
In the $EXP_{2F}$ state, two failed disks need to be replaced by the brand-new ones, while we assume that both disks are replaced simultaneously. 
A human error in the disk replacement process, regardless it happened on one or both disks, moves the array to the $DU_{FFH}$ state, while 
a successful disk replacement moves the array to the $EXP_{2FR}$ state. 
In the $EXP_{2FR}$ state, the array has two failed disks that are replaced with the brand-new ones, and the data of two failed disks should be recovered using other $n-2$ operating disks. 
In this state, two disks can be recovered simultaneously, or be recovered one after another. 
Both approaches take a minimum time twice the minimum time of one disk recovery, while the latter approach has reliability benefits, 
as after the recovery of the first disk the array moves to $OP_{1FR}$ state and stays a shorter time in the $EXP_{2FR}$ state. 
In the first approach, the array remains in the $EXP_{2FR}$ until the recovery of both disks.   
Hence, we take the latter approach in our simulations, as shown in Fig.~\ref{fig:raid6hr}.
Finally, $DU_{FFH}$ stands for the state in which user data is unavailable due to two disk failures and one human error in the array. In this state, by recovering from human error the array moves to the $EXP_{2FR}$ state. However, if the wrongly removed disk crashes, the array moves to the $DL_{TDF}$ state (triple disk failure). 

\subsection{Dependability of $RAID5$ Considering LSE}
\label{sec:raid5-lse}

\begin{figure}
\begin{centering}
\includegraphics[width=3.6in]{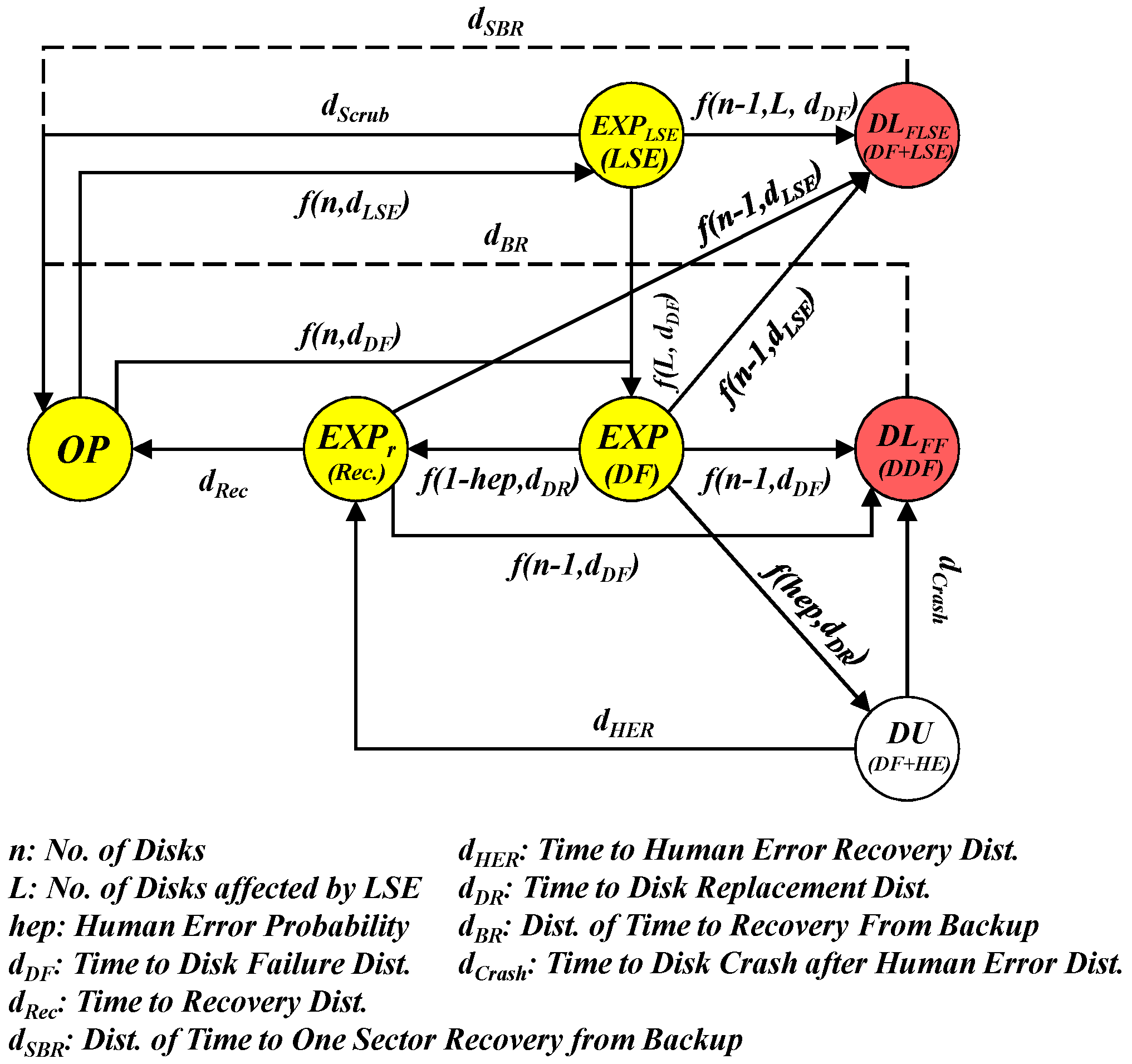}
\caption{State diagram of Monte Carlo simulation for assessing $RAID5$ DU/DL, considering LSE.}
\label{fig:raid5hrlse}
\par\end{centering}
\vspace{-0.3cm}
\end{figure}

The model presented in Fig.~\ref{fig:raid5hr} is extended to include LSE, as well as disk failures and IDRS, shown in Fig.~\ref{fig:raid5hrlse}. 
In the case a disk failure is followed by a human error, DU happens which can be assessed as described in Section~\ref{sec:analysis-no lse-no spare}. 
DL is another possible incidence when two consecutive disk failures or the combination of LSE and disk failure (on two different disks) happen. 

In the $OP$ state, all disks are operating with no LSE. 
An operational disk failure switches the array state to the $EXP$ state, while the time to transition from $OP$ to $EXP$ is a function of number of disks, $n$,  and the distribution of time to disk failure, $d_{DF}$. 
The array switches from $OP$ state to $EXP_{LSE}$ state when one or more LSEs happen. 
The LSE can be recovered by data scrubbing, while the time to scrub, with the distribution of $d_{Scrub}$ depends on the storage maintenance policies and can have a minimum value which 
depends on the array throughput~\cite{elerath-DSN-2007}.
A disk failure after an LSE on a different disk results in the loss of data which is damaged by LSE (state $DL_{FLSE}$). 
Elerath and Pecht~\cite{elerath-DSN-2007} take this incidence as DDF, while it has a different magnitude of data loss (and consequently a different 
recovery time in the case of survivable array) compared to DDF, resulting DL and DU overestimation.
The survivable sectors can be recovered from backups while the distribution of time to recover sectors from backups, $d_{SBR}$ is a function of 
the number of lost sectors, backup throughput, array throughput, and network speed\footnote{Here we can note a limitation of Markov model over 
Monte Carlo simulations; The Markov model cannot hold the number of LSEs, and consequently cannot accurately model the recovery time, as well as the magnitude of data loss. It mandates taking simplified assumptions in Markov models, such as assuming that only one sector is affected by LSE.}.  
In the $DL_{FLSE}$ state, by assuming $DOS(t)$ as the data survivability, the NOMDL and NOMDU imposed by DL incidence $i$ is evaluated respectively by Equation~\ref{equ:nomdl-dos} and Equation~\ref{equ:nomdu-dos}, while the \emph{Logical Size of Lost Data} is equal to the \emph{Size of Sectors Affected by LSE}.

A transition from $EXP_{LSE}$ state to $EXP$ state is possible when the only disk affected by LSE fails.
The time to failure of LSE affected disk can be different from operating disks, as it can have alternative causes such as 
\emph{Excessive Block Reallocation} and can be measured using field data~\cite{elerath-DSN-2007}, 
while it has no explicit rate and is included in $d_{DF}$~\cite{elerath-DSN-2007}.  
Hence, the time to transition from $EXP_{LSE}$ to $EXP$ also follows $d_{DF}$.
When one of $n-1$ LSE-free disks fails, the combination of LSE and disk failure moves the array from $EXP_{LSE}$ to $DL_{FLSE}$, where time 
to transition is a function of $n-1$ and $d_{DF}$.  
Note if more than one disk is affected by LSEs, a consequent failure of any disk moves the array to $DL_{FLSE}$ (hence, the time to transition is a function of $n$ rather than $n-1$) and there is no transition from $EXP_{LSE}$ to $EXP$\footnote{
This case also cannot be accurately modeled by Markov, as the Markov cannot recognize whether only one disk is affected by LSEs.}.
Hence, both transitions from $EXP_{LSE}$ to $EXP$ and $DL_{FLSE}$ is also a function of $L$, the number of disks affected by LSE.

In $EXP$ and $EXP_r$ states, the occurrence of LSE before the completion of disk replacement or disk recovery can move the array to $DL_{FLSE}$. 
However, Elerath and Pecht~\cite{elerath-DSN-2007} express that this transition has a low probability and ignore it.  

$DL_{FF}$ and $DU$ states are the same as $DL$ and $DU$ states in Fig.~\ref{fig:raid5hr}, while the imposed NOMDL and NOMDU can be assessed by
Equation~\ref{equ:nomdl-dos} and Equation~\ref{equ:nomdu-dos} in the case of $DL_{FF}$, 
and in the case of $DU$, NOMDU can be assessed by Equation~\ref{equ:nomdu}.

\subsection{Dependability of $RAID5$ With Automatic Fail-over Considering LSE}
\label{sec:raid5-spare-lse}
\begin{figure}
\begin{centering}
\includegraphics[width=3.2in]{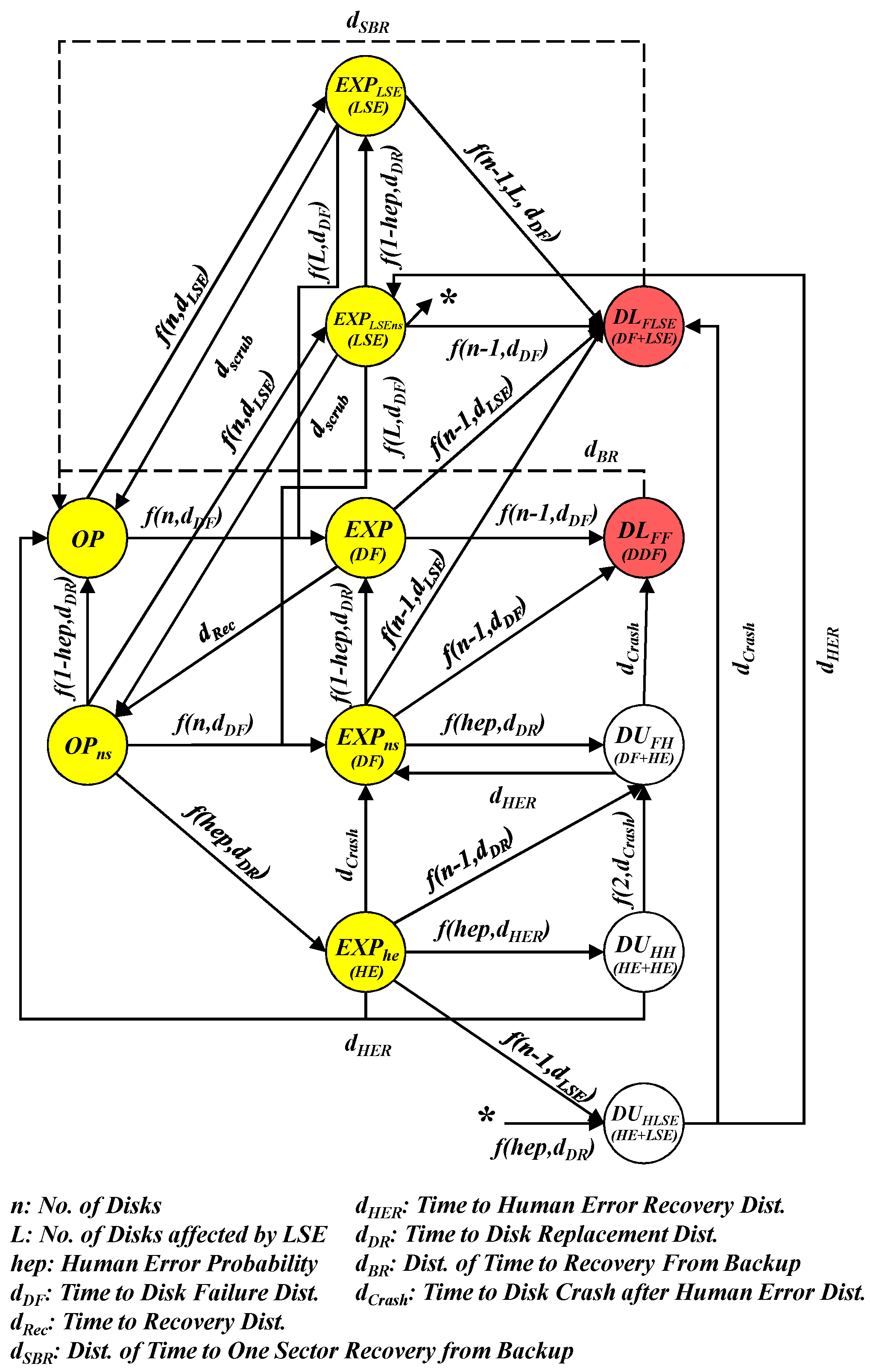}
\caption{State diagram of Monte Carlo simulation for $RAID5$ DU/DL with automatic fail-over, considering LSE.}
\label{fig:raid5-spare-lse}
\par\end{centering}
\vspace{-0.3cm}
\end{figure}

The final model presented in this section belongs to $RAID5$ array with hot spare disk, in which the delayed disk replacement policy is employed. 
In this policy, the disk replacement is performed after the completion of automatic recovery (to the spare disk), when the single point of failure is removed. 
Hence, this policy forbids the human error following disk failure which results in DU in the case of no spare (Section~\ref{sec:raid5-lse}).
The state diagram for obtaining DU/DL using Monte Carlo simulations is appeared in Fig.~\ref{fig:raid5-spare-lse}.

Upon a disk failure, the array moves from $OP$ to either $EXP$ and $DL_{FF}$ upon the first and second disk failures, respectively. 
In the $EXP$ state, the automatic recovery starts on the spare disk, by distribution $d_{DF}$, while the service agent has to forbid changing the failed 
disk with brand-new one, before the completion of recovery. 
After recovery, the array moves to $OP_{ns}$, where the array is operational but no spare disk is available. 
In this state, failed disk replacement can be performed by the service agent. 
The successful disk replacement moves the array back to the $OP$ state, while the human error moves the array to $EXP_{he}$ state. 
In the $EXP_{he}$ state, one operating disk is removed due to human error and the array is working with $n-1$ operating disks. 
In this state, another disk failure and an LSE moves the array to either $DU_{FH}$ and $DU_{HLSE}$ states, respectively, while a successive human error 
in disk replacement moves the array to $DU_{HH}$ states.
In the $DU_{FH}$ and $DU_{HH}$ states, the whole array is unavailable while 
in the $DU_{HLSE}$ state, only the sectors affected by LSE are unavailable and 
the \emph{Logical Size of Unavailable Data} is equal to the \emph{Size of Sectors Affected by LSE}.
 The imposed NOMDU by $DU_{FH}$, $DU_{HH}$, and $DU_{HLSE}$ incidences is obtained by Equation~\ref{equ:nomdu}.

In $OP_{ns}$ state, a disk failure and LSE moves the array to either $EXP_{ns}$ and $EXP_{LSEns}$ states, respectively. 
In $EXP_{ns}$ and $EXP_{LSEns}$ states the array has no spare, while successful replacement of the failed disk moves the array to either 
$EXP$ and $EXP_{LSE}$ states, respectively. 
Unsuccessful disk replacement in $EXP_{ns}$ and $EXP_{LSEns}$ states results in DU, moving the array to either $DU_{FH}$ and $DU_{HLSE}$ state, respectively.

\subsection{Monte Carlo Simulation}
\label{sec:monte carlo simulation}
\begin{figure*}
\begin{centering}
\includegraphics[width=0.8\textwidth]{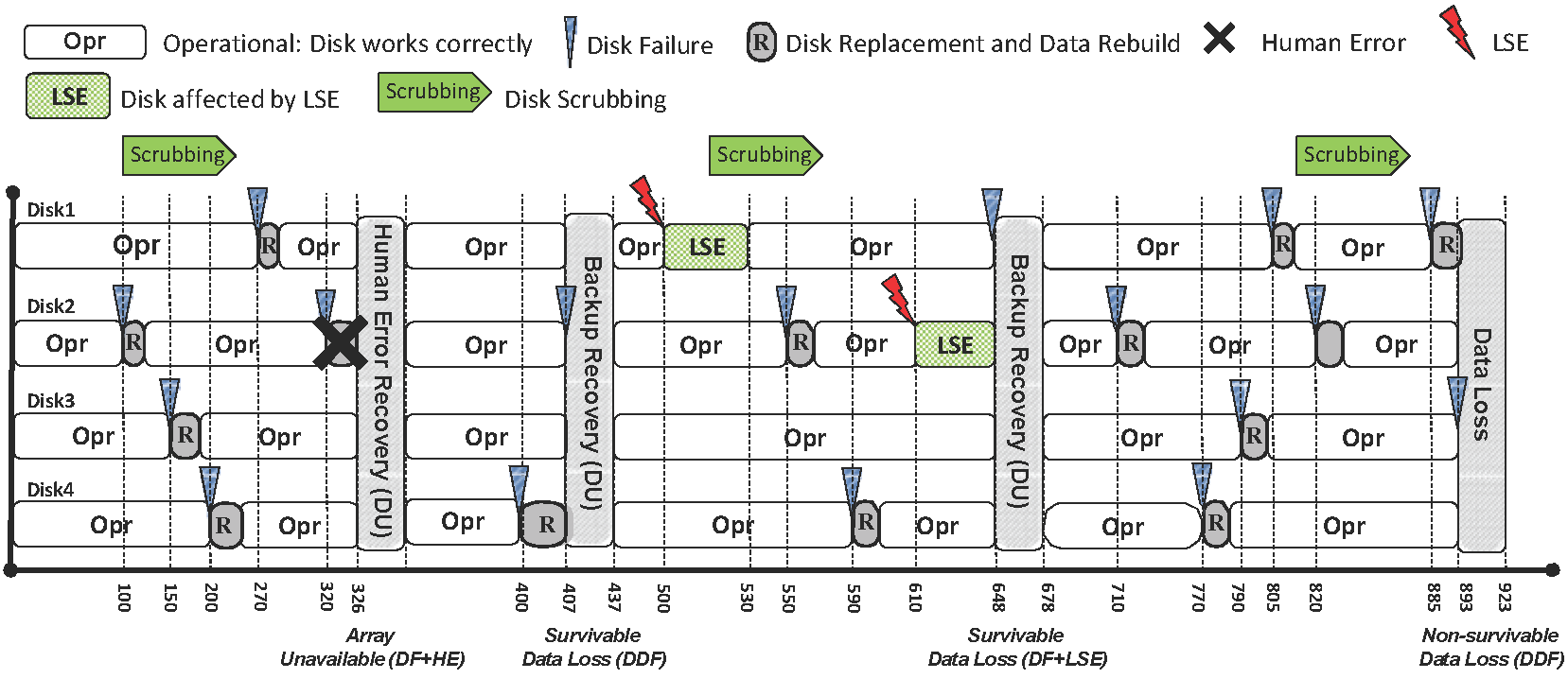}
\caption{MC Simulation to assess NOMDU and NOMDL of a $RAID5(3+1)$ Array in Presence of Human Errors}
\label{fig:raid5-simulation-events}
\par\end{centering}
\vspace{-0.3cm}
\end{figure*}

In the MC simulations, the disk failure and LSE incidences are generated by assuming the desired failure distributions such as \emph{Weibull} and exponential. 
After a disk failure occurrence, the recovery time is evaluated depending on the defined average recovery distribution.  
Fig.~\ref{fig:raid5-simulation-events} illustrates an example of the MC simulation for a $RAID5$ $(3+1)$ array. 
In case of DDF, i.e., two consecutive disk failures in the same array 
while the second failure is before the recovery of the first failure, a DL event happens (at $407$ and $893$ in Fig.~\ref{fig:raid5-simulation-events}), 
while the DL is recovered from backup if happened on survivable data (time $407$) or is permanently lost if happened on non-survivable data (time $893$).

In the case of single disk failure, the failed disk is replaced by a human agent. 
However, the occurrence of a human error in the disk replacement process, by the probability of $hep$, makes another working disk unavailable, resulting in the unavailability of the 
entire data array (time $326$).
The combination of LSE with disk failure and human error result in DL and DU, respectively. 
For example, at time $610$ an LSE happen on $disk2$, while the failure of $disk1$ at time $648$ results DL in the affected sectors, mandating the 
recovery of lost sectors from backup. 
Disk scrubbing is periodically performed on each disk and removes LSEs, while the exact time of removing each LSE is defined by considering a uniform distribution between start-time and end-time of scrubbing. 
For example, at time $500$ an LSE happen on $disk1$ that is removed by scrubbing at time $530$.
$NOMDU$ and $NOMDL$ is evaluated for each failure incidence, and is aggregated within mission time.

The error of MC simulations is inversely proportional to the root square of the number of iterations as shown in Equation~\ref{equ:A_Steady}. The number of iterations can be adjusted by the target accuracy (error) and the given confidence level.
Error of Monte Carlo simulation is obtained by Equation~\ref{equ:A_Steady}~\cite{lange1989robust}.

\begin{equation}
\label{equ:A_Steady}
\begin{centering}
Error_{Monte~Carlo}=\frac{\delta\times Z_{\alpha/2}}{\sqrt{n}}
\end{centering}
\end{equation} 

In Equation~\ref{equ:A_Steady}, $n$ is the number of iterations (in our case $n=number~of~simulated~arrays=1000$), $\delta$ is the standard deviation of the target values (NOMDU and NOMDL in our case), and
$Z_{\alpha/2}$ is the \emph{t-student} coefficient for a target confidence level~\cite{lange1989robust}.

\subsection{Monte Carlo Transitions}
\label{sec:monte-carlo-transitions}
The MC simulations can be applied to any failure and repair distribution, including exponential and Weibull. 
Elerath and Schindler~\cite{elerath2014beyond} consider a two-parameter Weibull distribution for time to disk failures, LSEs, recovery of disk failures, and scrubbing, and show that this distribution better corroborates the field data, compared to the exponential distribution.
This distribution assumes the probability density
function as shown in Equation~\ref{equ:weibull} where $t$ is time, $\eta$ is the characteristic life,  
$\gamma$ is location parameter, and $\beta$ is the shape parameter~\cite{nelson2005applied}.

\begin{equation}
\label{equ:weibull}
f(t)~={\left ( \frac{\beta }{\eta } \right )\left ( \frac{t-\gamma}{\eta } \right )^{\beta-1}exp\left [ -\left ( \frac{t-\gamma}{\eta } \right )^{\beta} \right ]}
\end{equation}

We use the base parameters obtained from field data by Elerath and Schindler~\cite{elerath2014beyond}, as shown in Table~\ref{tab:weibull-parameters}. 
Note as Elerath and Schindler use two-parameter Weibull, we need to consider $\gamma = 0$ when applying Table~\ref{tab:weibull-parameters} parameters to Equation~\ref{equ:weibull}.

\begin{table}[]
\centering
\caption{Disk Failure, Disk Failure Reconstruct, LSE, and Scrubbing Weibull distribution parameters for three disk models from 10,000 storage systems in the field~\cite{elerath2014beyond}. Disk A and Disk B are 1TB near-line SATA models and have been in the field for average 3 years, and Disk C is an enterprise-class FC 288GB model and has been in the field for average 5 years.}
\label{tab:weibull-parameters}
\begin{adjustbox}{width=0.5\textwidth,totalheight=\textheight,keepaspectratio}
\begin{tabular}{|c|c|c|c|c|c|c|c|c|}
\hline
\multirow{2}{*}{Disk Model} & \multicolumn{2}{c|}{Disk Failure ($d_{DF}$)} & \multicolumn{2}{c|}{Recovery ($d_{Rec}$)} & \multicolumn{2}{c|}{LSE ($d_{LSE}$)} & \multicolumn{2}{c|}{Scrubbing ($d_{Scrub}$)} \\ \cline{2-9} 
                            & $\eta_{DF}$          & $\beta_{DF}$          & $\eta_{Rec}$        & $\beta_{Rec}$       & $\eta_{LSE}$     & $\beta_{LSE}$     & $\eta_{Scrub}$       & $\beta_{Scrub}$       \\ \hline
SATA Disk A                 & 302,016              & 1.13                  & 22.7                & 1.65                & 12,325           & 1                 & 186                  & 1                     \\ \hline
SATA Disk B                 & 4,833,522            & 0.576                 & 20.25               & 1.15                & 42,857           & 1                 & 160                  & 0.97                  \\ \hline
FC/SCSI Disk C              & 1,058,364            & 0.721                 & 6.75                & 1.4                 & 50,254           & 1                 & 124                  & 2.1                   \\ \hline
\end{tabular}
\end{adjustbox}
\end{table}

For disk replacement and human error recovery, we also cannot assume a constant rate (exponential distribution), as by this assumption the probability of 
disk replacement and human error recovery in any time interval with the equal size is the same, which is not realistic. Hence, we also use Weibull distribution for disk replacement and human error recovery. 
The time to disk replacement, with the distribution of $d_{DR}$, has no minimum value, as the human agent can change the failed disk immediately after its failure. 
Hence, we consider minimum time of 0 hours for the location parameter ($\gamma = 0$).
We consider shape parameter ($\beta$) of 2 to have a right-skewed distribution, similar to the disk restore distribution.
We consider the characteristic life of half an hour ($\eta=0.5$), obtained from the storage service logs of \emph{Sharif University of Technology}~\cite{sharifuni} datacenter, as a typical expected time for the failed disk replacement.

Time to recognize and recover the human error is denoted by $d_{HER}$.
As the human error can be recognized and recovered immediately, 
we consider minimum time of 0 hours for the location parameter ($\gamma = 0$). 
The shape parameter of 2 is considered to have a right-skewed distribution, and the characteristic life of one hour ($\eta=1$) is considered regarding  
our storage service logs and interviews with datacenter technicians.  
Time to crash the wrongly replaced disk is generated by considering the shape parameter 1.4, and the characteristic life of one year ($\eta=8760$), 
obtained by our storage service logs. 
The location parameter is 0 ($\gamma = 0$), as the wrongly replaced disk can be immediately thrown away.
The Weibull parameters corresponding to disk replacement and human error is appeared in Table~\ref{tab:human-error-parameters}.
 
\begin{table}[]
\centering
\caption{Human error parameters from field data and interview with datacenter technicians.}
\label{tab:human-error-parameters}
\begin{adjustbox}{width=0.5\textwidth,totalheight=\textheight,keepaspectratio}
\begin{tabular}{|c|c|c|c|c|c|c|c|c|}
\hline
\multicolumn{2}{|c|}{Disk Replacement ($d_{DR}$)} & \multicolumn{2}{c|}{Human Error Recovery ($d_{HER}$)} & \multicolumn{2}{c|}{Crash Wrongly Replaced Disk ($d_{Crash}$)}  \\ \hline
$\eta_{DR}$       & $\beta_{DR}$      &  $\eta_{HER}$       & $\beta_{HER}$      & $\eta_{Crash}$       & $\beta_{Crash}$  \\ \hline
0.5            & 2      & 1           & 2           & 8760          & 1.4         \\ \hline
\end{tabular}
\end{adjustbox}
\vspace{-0.3cm}
\end{table}

Time to backup recovery in the case of DL in survivable storage, $d_{BR}$, can also be characterized by a three-parameter Weibull distribution. 
In the case of DDF, the data of two failed disks is obtained from the backup. 
An alternative is to obtain the data of the first failed disk from the backup, afterwards, reconstruct the second failed disk using the XOR of $n-1$ operating disks of the array. 
Assuming a network connection of $1Gbps$ between the storage and backup, and considering the array has eight 500GB SATA disks with $50MBps$ speed, 
obtaining the data of failed disk from backup takes 10 hours. 
Considering the disks are connected to a $1.5Gbps$ data bus,  
it also takes 10.4 hours to reconstruct the failed disk using the XOR of $n-1$ operating disks of the array~\cite{elerath-DSN-2007}. 
Hence, a minimum time of 20 hours is required to recover a DDF from backup ($\gamma = 20$).
We consider twice of the minimum recovery time as the characteristic life ($\eta=40$), and consider the shape parameter of 2, to have a right skewed distribution. 
In the case of DL in disk sectors, caused by LSE, the distribution of recovery time, $d_{SBR}$, depends on the size of lost sectors. 
As one sector typically has an small size of $4KB$, the minimum backup recovery time depends on the minimum disk response 
time and the network delay, while we consider one millisecond for minimum sector recovery from the backup ($\gamma = 2.7 \times 10^{-7}$), 
two millisecond for the characteristic life ($\eta = 5.5 \times 10^{-7}$), and the shape parameter of 2 ($\beta = 2$) to have a right skewed distribution.       
The Weibull parameters corresponding to $d_{BR}$ and $d_{SBR}$ are appeared in Table~\ref{tab:dl-recovery-parameters}.

\begin{table}[]
\centering
\caption{Data loss recovery parameters from field data and interview with datacenter technicians.}
\label{tab:dl-recovery-parameters}
\begin{adjustbox}{width=0.4\textwidth,totalheight=\textheight,keepaspectratio}
\begin{tabular}{|c|c|c|c|c|c|}
\hline
\multicolumn{3}{|c|}{Backup Recovery ($d_{BR}$)} & \multicolumn{3}{c|}{Sector Backup Recovery ($d_{SBR}$)}     \\ \hline
$\gamma_{BR}$   & $\eta_{BR}$   & $\beta_{BR}$   & $\gamma_{SBR}$       & $\eta_{SBR}$         & $\beta_{SBR}$ \\ \hline
20              & 40            & 2              & $2.7 \times 10^{-7}$ & $5.5 \times 10^{-7}$ & 2             \\ \hline
\end{tabular}
\end{adjustbox}
\vspace{-0.3cm}
\end{table}

\subsection{\hl{Applying Proposed Model to General Erasure Codes}}
\label{sec:general-erasure-codes}
\hl{In the previous subsection, we discussed the effect of human errors in $RAID5$ and $RAID6$ arrays and clarified how we use Monte Carlo simulations to obtain NOMDL and NOMDU for a specific array architecture by considering disk failures, LSEs, and human errors.
However, both $RAID5$ and $RAID6$ schemes are in the category of \emph{Maximum Distance Separable} (MDS) codes. Many alternatives of MDS codes are proposed in the recent years to cope with failure types observed in HDD and SSD arrays. Hence, it is of great importance that our proposed Monte Carlo framework cope with MDS codes in general case. }

\begin{figure}
    \centering
  
        \includegraphics[width=0.3\textwidth]{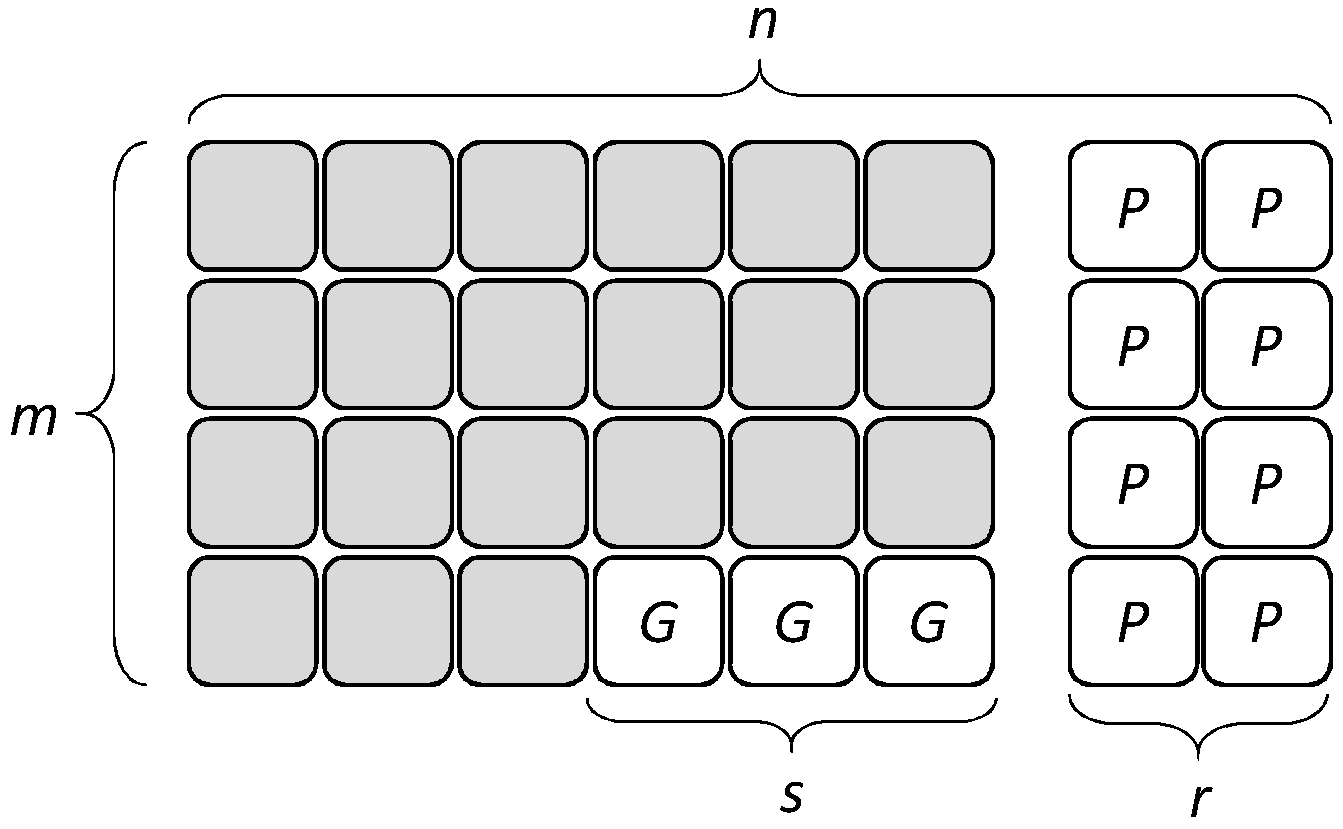}
    \caption{\hl{Scheme of Partial-MDS codes}}
    \label{fig:general-erasure-codes}
    \vspace{-0.3cm}
\end{figure}

\hl{MDS codes, proposed in $70^{th}$, offer the maximum possible hamming distance (hence, the maximum correction capability) while being separable, and have many alternatives such as Parity codes, Reed-Solomon codes~\cite{macwilliams1977theory,plank1997tutorial}, or array codes, such as EVENODD~\cite{Blaum-TC-95}, RDP~\cite{corbett2004row}, X-codes~\cite{xu1999x}, B-codes~\cite{xu1999low}, HVD codes~\cite{kishani2011hvd}, Liberation codes~\cite{Plank-2008-FAST}, STAIR codes~\cite{li2014stair}, Sector-Disk Codes~\cite{plank2014sector}, and Partial-MDS codes~\cite{blaum2013partial}. $RAID5$ and $RAID6$ configurations are also in the MDS category by keeping respectively one and two redundant parities to respectively cope with one and two device failures in a disk array. In a $RAID5$ configuration, a row-wise code-word (Parity code) is stored in a redundant data chunk (or in general, data symbol). The redundant data alongside the actual data constitutes a data stripe.
Blaum et al.~\cite{blaum2013partial} propose a Partial-MDS code that uses the conventional row-wise parity alongside a new concept of \emph{Global Parity} to cope with the combination of both device failures and symbol failures. In general, we have a linear $[mn, m(n-r)-s]$ code where $m$ is the number of rows per stripe (code-word), $n$ is the number devices in a stripe (including redundant devices), $r$ is the number of redundant devices, and $s$ is the number of global parities, as shown in Fig~\ref{fig:general-erasure-codes}.
}

\hl{In Partial-MDS codes (Fig.~\ref{fig:general-erasure-codes}), the $P$ (Parity) symbols are taken row-wise, while $G$ (Global Parity) symbols are taken globally from all array members. Blaum et al.~\cite{blaum2013partial}, Plank and Blaum~\cite{plank2014sector}, and Li and Lee~\cite{li2014stair} propose different approaches for encoding/decoding of Global parities by different complexities and I/O overhead. This code can cope with $r$ device failures and $s$ symbol failures in each code-word. We can put $RAID5$ in the category of Partial-MDS codes by considering $r=1$ and $s=0$. Similarly, we can put $RAID6$ in the category of Partial-MDS codes by considering $r=2$ and $s=0$. Briefly, we use the term $PMDS(m, n, r, s)$ to refer to a Partial-MDS code with $m$ rows, $n$ devices, $r$ row parities, and $s$ global parities.}

\subsubsection{\hl{Overheads of General Erasure Codes}}
\hl{Depending on the number of row parities and global parities, PMDS codes come with different I/O overhead, computational complexity, and \emph{Effective Replication Factor} (ERF\footnote{ERF stands for the ratio of storage physical capacity over storage logical (useful) capacity.}), while the computational complexity and ERF is analyzed in the previous work~\cite{li2014stair,plank2014sector,blaum2013partial}.
In general, ERF of $PMDS(m, n, r, s)$ is calculated by Equation~\ref{equ:erf}.
}

\begin{equation}
\label{equ:erf}
\begin{split}
ERF[PMDS(m,n,r,s)] = \frac{m \times n }{m \times (n-r) - s}
\end{split}
\end{equation}

\subsubsection{\hl{Dependability Analysis of General Erasure Codes}}
\hl{In the general case, we can consider four failure types for a disk array:}
\begin{itemize}
\item
\hl{\textbf{\emph{Array Data Loss (ADL):}} This failure is similar to what we previously called DDF in the case of $RAID5$, and TDF in the case of $RAID6$, in which the whole array is lost. }
\item
\hl{\textbf{\emph{Stripe Data Loss (SDL):}} is named after the failure case in which one or multiple stripes of disk array is lost.}
\item
\hl{\textbf{\emph{Array Data Unavailability (ADU):}} is named after the failure case in which the whole array is unavailable due to human errors (IDRS).}
\item
\hl{\textbf{\emph{Stripe Data Unavailability (SDU):}} is named after the failure case in which one or multiple stripes of disk array is unavailable due to human errors (IDRS).}
\end{itemize}

\hl{Consider employing $PDMS(m,n,r,s)$ in a disk array as detailed in Table~\ref{tab:pmds-array-assumptions}. By considering the definitions shown in Table~\ref{tab:pmds-reliability-definitions}, the conditions of $ADL$, $SDL$, $ADU$, and $SDU$ failures are summarized in Table~\ref{tab:pmds-failure-conditions}.
$ADL$ happens in a very simple condition, when the number of failed devices ($DF$) surpasses $r$ (the number of redundant devices).
$SDL$ happens when $ADL$ condition is not satisfied, but there exist at least one stripe in which the number of LSEs surpasses the maximum correctable LSEs.
$ADU$ happens when $ADL$ condition is not satisfied, but the aggregation of failed devices ($DF$) and unavailable devices by human error ($HE$) surpasses $r$.
Note it is possible that both $ADU$ and $SDL$ conditions are satisfied in some cases, when the whole array is unavailable while some of array stripes is lost.
Finally, $SDU$ happens when $ADU$ and $ADL$ conditions are not satisfied and at least one stripe exists in which the number of LSEs does not surpass the maximum correctable LSEs, but its data is unavailable due to human error.
Note it is possible that both $SDU$ and $SDL$ conditions are satisfied in some cases, when the array has at least one unavailable stripe and at least one lost stripe.
}

\hl{We conduct Monte Carlo simulations using the framework described in Section~\ref{sec:monte carlo simulation} and check the failure conditions appeared in Table~\ref{tab:pmds-failure-conditions} to recognize $ADL$, $SDL$, $ADU$, and $SDU$ failure cases. For each failure case, we record the size of lost data (in the case of $ADL$ and $SDL$) or size of unavailable data and unavailability duration (in the case of $ADU$ and $SDU$), and finally calculate NOMDU and NOMDL at the end of simulation using Equation~\ref{equ:nomdu} through Equation~\ref{equ:nomdl-dos}.  
}

\begin{table}
\centering
    \caption{\hl{Assumptions of employing $PMDS(m,n,r,s)$.}}
\label{tab:pmds-array-assumptions}
\begin{adjustbox}{width=0.5\textwidth,totalheight=\textheight,keepaspectratio}
\begin{tabular}{|c|c|}
\hline
\multirow{4}{*}{$PMDS(m,n,r,s)$ } & $m$: number of rows per stripe (codeword)   \\ \cline{2-2} 
            &  $n$: number of devices per array (number of chunks per stripe)\\ \cline{2-2} 
                  & $r$: number of row parities (redundant devices) \\ \cline{2-2} 
                  & $s$: number of global parities (redundant sectors) per stripe \\ \hline
\end{tabular}
 \end{adjustbox}   
\vspace{-0.3cm}
\end{table}

\begin{table*}
\centering
    \caption{\hl{Definitions for assessing dependability of $PMDS(m,n,r,s)$}}
    \vspace{-0.2cm}
\label{tab:pmds-reliability-definitions}
\begin{adjustbox}{width=0.8\textwidth,totalheight=\textheight,keepaspectratio}
\begin{tabular}{|c|}
\hline
  $V$: set of array stripes \\ \hline
  $DF$: number of failed devices \\ \hline
  $HE$: number of unavailable (wrongly removed) devices due to human error (IDRS) \\ \hline
  $NUM_{LSE}(i,v)$: number of LSEs (lost sectors) in chunk (device)  $i$ of stripe $v$ (0 for failed devices) \\ \hline
  $MAX(i,v)$: device number (excluding failed devices) having $i^{th}$  maximum⁡number of LSEs in stripe $v$ \\ \hline
  $MAXOP(i,v)$: operational device number (excluding unavailable and failed devices) having $i^{th}$  maximum⁡number of LSEs in stripe $v$ \\ \hline
  $OP(i)$: 1, device $i$ is operational (neither unavailable nor failed), 0, otherwise \\ \hline
\end{tabular}
 \end{adjustbox}   

\end{table*}

\begin{table*}
\centering
    \caption{\hl{Failure conditions in $PMDS(m,n,r,s)$}}
    \vspace{-0.2cm}
\label{tab:pmds-failure-conditions}
\begin{adjustbox}{width=1\textwidth,totalheight=\textheight,keepaspectratio}
\begin{tabular}{|c|c|}
\hline
\multicolumn{2}{|c|}{\cellcolor[HTML]{C0C0C0} \textbf{Failure Conditions in $PMDS(m,n,r,s)$}} \\ \hline
\textbf{\emph{ADL}} & $r<DF$ \\ \hline
\textbf{\emph{SDL}} & $(DF \leq r) \wedge (\exists v \in V [s + \sum_{i=1}^{r-DF} NUM_{LSE}(MAX(i,v),v) < \sum_{i=1}^{n} NUM_{LSE}(i,v)]) $ \\ \hline
\textbf{\emph{ADU}} & $(DF \leq r) \wedge ( r<DF+HE$) \\ \hline
\textbf{\emph{SDU}} &  $(0<HE) \wedge (DF+HE \leq r) \wedge (\exists v \in V [(\sum_{i=1}^{n}NUM_{LSE}(i,v) \leq s + \sum_{i=1}^{r-DF}NUM_{LSE}(MAX(i,v),v)) \wedge (s + \sum_{i=1}^{r-DF-HE}NUM_{LSE}(MAXOP(i,v),v) < \sum_{i=1}^{n}NUM_{LSE}(i,v) \times OP(i))]) $\\ \hline
\end{tabular}
 \end{adjustbox}   
\vspace{-0.3cm}
\end{table*}

\section{Simulation Results}
\label{sec:results}

\subsection{Experimental Setup}
Monte Carlo simulations are conducted for 1000 arrays of $RAID5(7+1)$ and the Weibull parameters appeared in Section~\ref{sec:monte-carlo-transitions} (Table~\ref{tab:weibull-parameters}, 
Table~\ref{tab:human-error-parameters}, and Table~\ref{tab:dl-recovery-parameters}).  
Each experiment simulates 10 years (87600 hours) of mission time. 
The Monte Carlo simulator is implemented from scratch in C++ with respect to the logic represented in Section~\ref{sec:monte carlo simulation}. 
The results of this section are obtained for a \emph{non-survivable storage system} (the definition of \emph{survivable storage systems} and \emph{non-survivable storage systems} is clarified in Section~\ref{sec:analysis-no lse-no spare}), hence, the recovery from DL states is not possible (in Fig.~\ref{fig:raid5hr}, Fig.~\ref{fig:raid6hr}, Fig.~\ref{fig:raid5hrlse}, and Fig~\ref{fig:raid5-spare-lse}, transition from $DL_{FLSE}$, $DL_{FF}$, and $DL_{TDF}$ states to $OP$ state, appeared in dashed-line, is impossible). In this regard, NOMDU and NOMDL are obtained respectively by Equation~\ref{equ:nomdu} and Equation~\ref{equ:nomdl}.

\subsection{Validating Monte Carlo Implementation}
\begin{figure}
    \centering
  
        \includegraphics[width=0.4\textwidth]{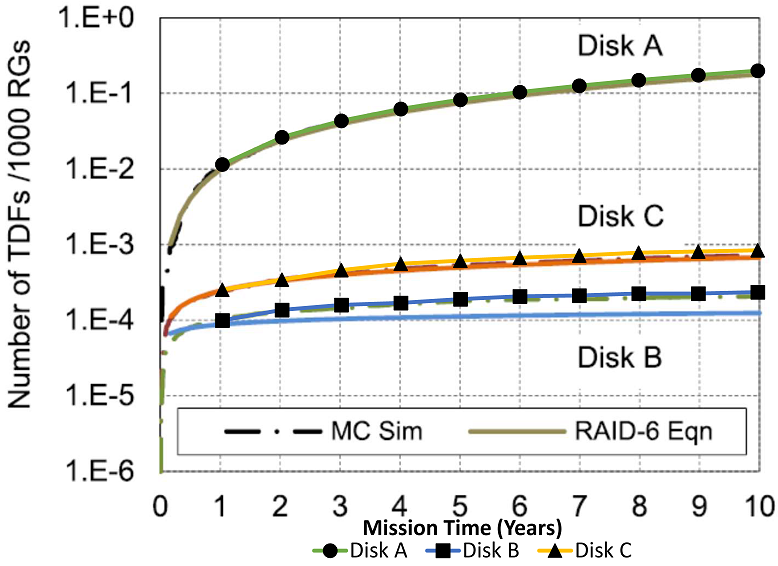}
    \caption{Monte Carlo simulation results for 10-years mission time, drawn on the results by Elerath and Schindler~\cite{elerath2014beyond}, for 1000 $RAID6(14+2)$ arrays of Disk A, Disk B, and Disk C.}
    \vspace{-0.6cm}
    \label{fig:compare-raid6-elerath}
\end{figure}

This is the first attempt of modeling the effect of human errors \hl{in data storage systems}. Hence, \hl{to validate the Monte Carlo implementation}, we compare the TDF within mission time obtained by our Monte Carlo implementation considering\emph{ no human errors}, with the \hl{Monte Carlo} results obtained by Elerath and Schindler~\cite{elerath2014beyond} for $RAID6$ array. In this comparison, we conduct the experiments for 1000 $RAID6(14 + 2)$ array groups and consider all data loss events, including DF+LSE+LSE, DF+DF+LSE, and DF+DF+DF, as TDF (Elerath and Schindler~\cite{elerath2014beyond} follow the same approach and consider all possible combinations of disk failure and LSE that result in data loss as TDF). In Fig.~\ref{fig:compare-raid6-elerath}, our simulation results for 10-years mission time is drawn versus the results by Elerath and Schindler~\cite{elerath2014beyond} for Disk A, Disk B, and Disk C models (considering the parameters appeared in Table~\ref{tab:weibull-parameters}). As Fig.~\ref{fig:compare-raid6-elerath} shows, our Monte Carlo simulations report slightly higher TDF values compared to previous work (on average 11\%).

\begin{figure}
    \centering
  
        \includegraphics[width=0.45\textwidth]{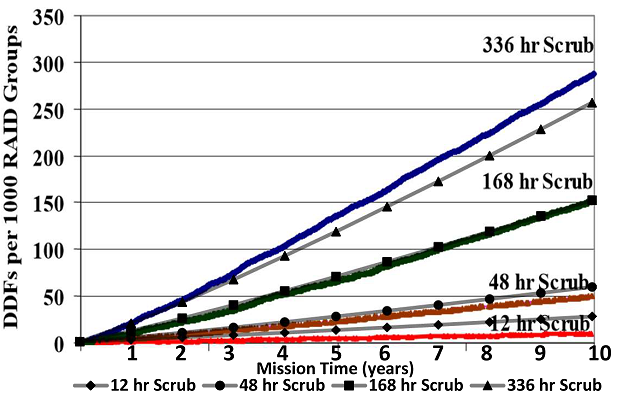}
    \caption{Monte Carlo simulation results for 10 years mission time, drawn on the results by Elerath and Pecht~\cite{Elerath-2009-TC,elerath-DSN-2007}, for different time to scrub ($\eta_{Scrub}$). The simulations 
    are conducted by the same basic parameters as Elerath and Pecht~\cite{Elerath-2009-TC,elerath-DSN-2007}: $\gamma_{DF}=0$, $\eta_{DF}=461386$, $\beta_{DF}=1.12$, $\gamma_{Rec}=6$, $\eta_{Rec}=12$, $\beta_{Rec}=2$, $\gamma_{LSE}=0$, $\eta_{LSE}=9259$, $\beta_{LSE}=1$, $\gamma_{Scrub}=6$, $\eta_{Scrub}=168$, $\beta_{Scrub}=3$.} 
    \label{fig:compare-elerath}
    \vspace{-0.3cm}
\end{figure}

We also compare the DDF within mission time obtained by our Monte Carlo implementation considering\emph{ no human errors}, with the results obtained by Elerath and Pecht~\cite{Elerath-2009-TC,elerath-DSN-2007} for $RAID5$ array. In this comparison, we conduct the experiments for 1000 $RAID5(7 + 1)$ groups and consider both LSE+DF and DF+DF incidences as DDF (Elerath and Pecht~\cite{Elerath-2009-TC,elerath-DSN-2007} follow the same approach and consider all possible combinations of disk failure and LSE that result in data loss as DDF). Hence, in the state diagram of Fig.~\ref{fig:raid5hrlse}, transition to both $DL_{FLSE}$ and $DL_{FF}$ states  is considered as DDF incidence. Table~\ref{tab:monte-carlo-comparison} compares the number of DDFs reported by Elerath and Pecht~\cite{Elerath-2009-TC,elerath-DSN-2007} with the results of our simulation for the first year of mission time. In Fig.~\ref{fig:compare-elerath}, our simulation results for 10-years mission time is drawn versus the results by Elerath and Pecht~\cite{Elerath-2009-TC,elerath-DSN-2007}. As the figure shows, for $\eta_{Scrub}=12$, 48, and 168  hours, our Monte Carlo simulations report greater number of DDFs, while for $\eta_{Scrub}=336$ hours, the model of Elerath and Pecht predicts greater number of DDFs. In summary, the difference of our Monte Carlo simulation results with the results by Elerath and Pecht is 56\%, 13\%, 1.3\%, and 9\%, respectively for $\eta_{Scrub}= 12$, 48, 168, and 336 hours.


\begin{table}
\centering
    \caption{Comparing our Monte Carlo implementation results with Elerath and Pecht~\cite{elerath-DSN-2007} in the first year of mission time for different time to scrub, in terms of \emph{number of DDF incidences}.}
\label{tab:monte-carlo-comparison}
\begin{adjustbox}{width=0.45\textwidth,totalheight=\textheight,keepaspectratio}
    \begin{tabular}{ | c | c | c |}
    \hline
	Time to Scrub & DDF by our Implementation & DDF by Elerath and Pecht~\cite{Elerath-2009-TC,elerath-DSN-2007} \\ \hline
	$\eta = 336$ hours & 20 & 21 \\ \hline
	$\eta = 168$ hours & 12 & 11 \\ \hline
	$\eta = 48$ hours & 5 & 5 \\ \hline
	$\eta = 12$ hours & 2 & 1 \\ \hline   
    \end{tabular}
 \end{adjustbox}   

\vspace{-0.3cm}
\end{table}

\subsection{Effect of Human Error in Non-survivable Storage System}

\begin{figure}
    \centering
	\subfigure[NOMDU]
	{  
        \includegraphics[width=0.35\textwidth]{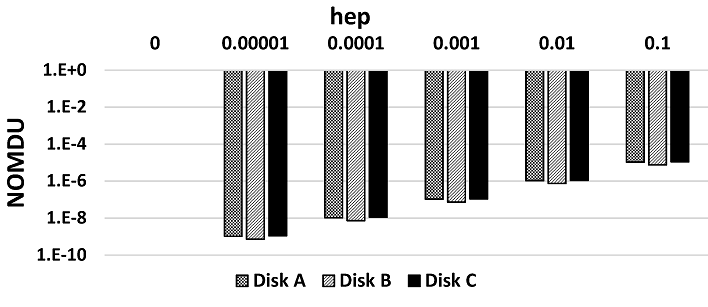}
        \label{fig:nomdu}
        }
    \subfigure[NOMDL-DDF]
    {
    	\includegraphics[width=0.35\textwidth]{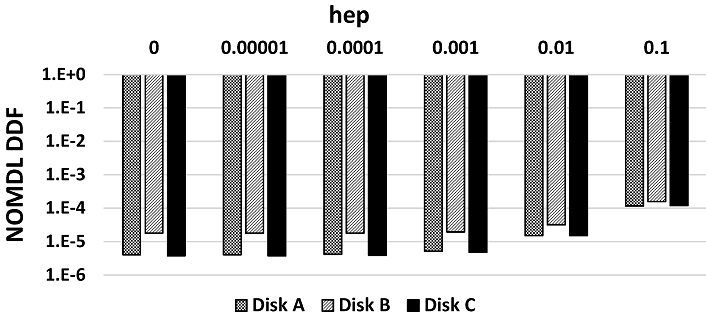}
    	\label{fig:nomdl-ddf}
    	}
  	\subfigure[NOMDL-DF+LSE]
  	{
  		\includegraphics[width=0.35\textwidth]{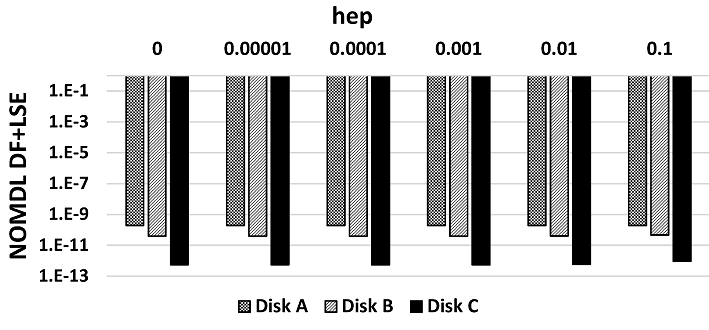}
  		\label{fig:nomdl-dflse}
  		}
    \caption{NOMDU and NOMDL caused by human errors for three different disk types (Table~\ref{tab:weibull-parameters}) and different \emph{hep}. The experiments are conducted for 1000 $RAID5(7+1)$ arrays. We differentiate NOMDL caused by DDF and LSE+DF, respectively appeared in sub-figures b and c.}
    \label{fig:nomdu-nomdl}
    \vspace{-0.3cm}
\end{figure}

Fig.~\ref{fig:nomdu-nomdl} reports NOMDU and NOMDL for $RAID5$ array, obtained by the model appeared in Fig.~\ref{fig:raid5hrlse}.
The experiments are conducted for $1000$ $RAID5(7+1)$ arrays of Disk A, Disk B, and Disk C (Table~\ref{tab:weibull-parameters}). 
We differentiate NOMDL caused by DDF and LSE+DF, respectively appeared in Fig.~\ref{fig:nomdl-ddf} and Fig.~\ref{fig:nomdl-dflse}.
Fig.~\ref{fig:nomdu} shows that by increasing $hep$ by one order of magnitude, NOMDU almost increases by one order of magnitude.
Meanwhile, increasing $hep$ has less impact on NOMDL caused by DDF, and negligible impact on NOMDL caused by DF+LSE. 
By increasing $hep$ from 0 to 0.001, the increase of both NOMDL caused by DDF and NOMDL caused by DF+LSE is negligible for all disk types.
By increasing $hep$ from 0 to 0.01 and 0.1, NOMDL caused by DF+LSE increases respectively by 1.0002x and 1.002x in arrays of disk A,  1.01x and 1.2x in arrays of disk B, and 1.07x and 1.8x in arrays of disk C. 
By increasing $hep$ from 0 to 0.01 and 0.1, NOMDL caused by DDF increases respectively by 4.7x and 38x in arrays of disk A,  2x and 10x in arrays of disk B, and 5.3x and 44x in arrays of disk C. 
We can conclude that human error increases NOMDU by one order of magnitude, while it has no impact on 
NOMDL when $hep$ is below 0.001, and this observation is almost regardless of disk type.
However, when $hep$ reaches 0.01 and beyond, it dramatically increases DL within mission time.

Another important observation is that NOMDL caused by LSE is five orders of magnitude smaller than NOMDL caused by DDF, while our simulation results show that LSE causes more than 90\% of all DL incidences. We can explain this observation by different magnitudes of data loss in DDF and DF+LSE incidences. While DDF makes the whole array lost, DF+LSE results in data loss of one or multiple stripes. This observation concludes that the approach proposed by Elerath and Pecht~\cite{elerath-DSN-2007,Elerath-2009-TC} in taking both DDF and DF+LSE the same will result in serious DL overestimation.


\subsection{Availability Comparison of RAID Configurations with Equivalent Usable Capacity}

In this section we investigate whether human errors can change our conventional assumptions about the dependability of different RAID configurations.
To this end, we compare NOMDL and NOMDU of 
$RAID5(3+1)$, $RAID5(7+1)$, and $RAID1(1+1)$ configurations,  
considering equivalent usable (logical) capacity.

\subsubsection{Applying the $RAID5$ dependability Models to $RAID1$}
$RAID1$ system is implemented by mirroring the disk data in a redundant disk. 
Hence, it can be modeled as a one-failure tolerant system.
Similar to $RAID5$, the data is lost in the case of DDF and disk failure combined with LSE, 
and the data is unavailable in the case of human error in disk failure recovery process.  
As such, the DU and DL is evaluated by the models presented in Section~\ref{sec:raid5-lse} and Section~\ref{sec:raid5-spare-lse}, 
by considering $n=2$.

\begin{figure}
    \centering
  	\subfigure[NOMDU]
  	{
        \includegraphics[width=0.35\textwidth]{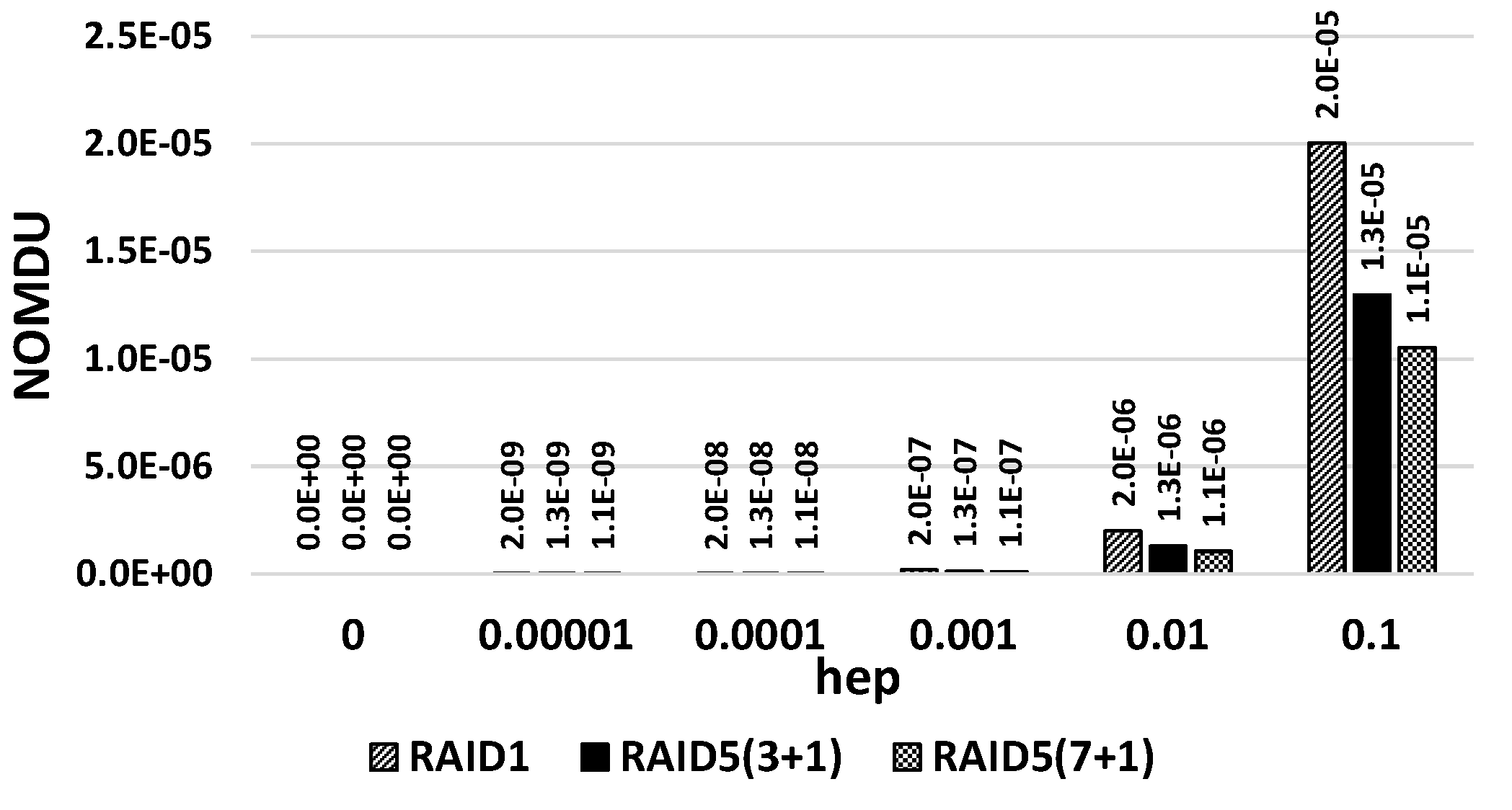}
        \label{fig:RAID-comparison-nomdu}
        }
    \subfigure[NOMDL-DDF]
    {
    	\includegraphics[width=0.35\textwidth]{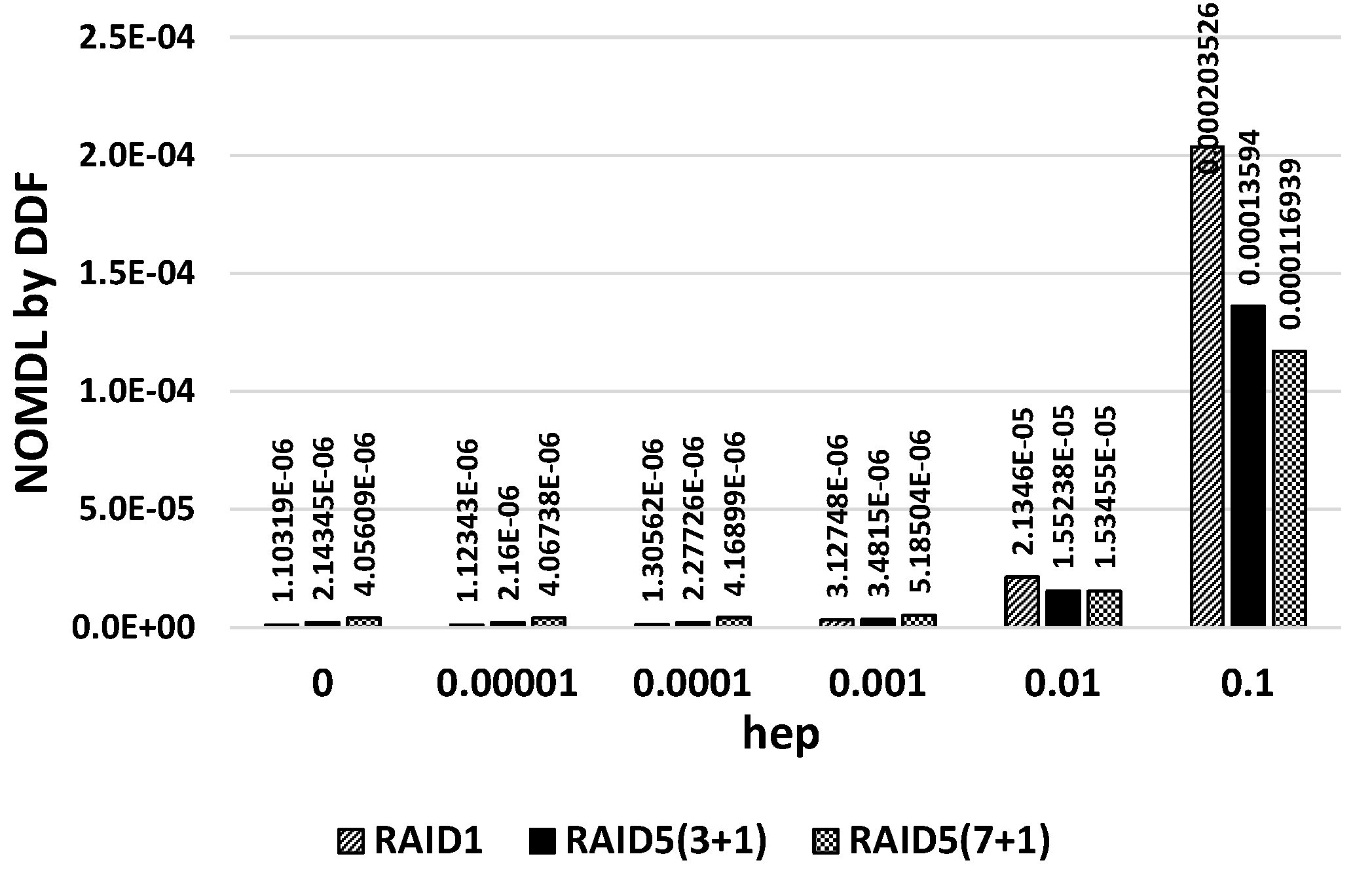}
    	\label{fig:RAID-comparison-nomdl-ddf}
    	}
    \subfigure[NOMDL-DF+LSE]
    {
    	\includegraphics[width=0.35\textwidth]{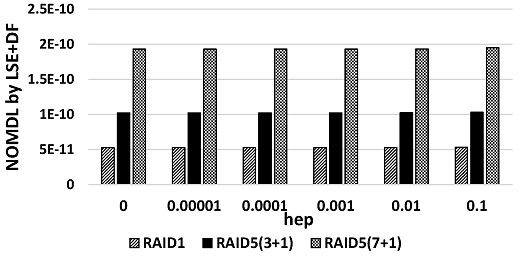}
    	\label{fig:RAID-comparison-nomdl-lsedf}
    	}
    \caption{NOMDU and NOMDL caused by human errors for different RAID configurations with equivalent usable capacity. The experiments are conducted for $21000$ $RAID1$ arrays, $7000$ $RAID5(3+1)$ arrays, and $3000$ $RAID5(7+1)$ arrays of Disk A (Table~\ref{tab:weibull-parameters}). We differentiate NOMDL caused by DDF and LSE+DF, respectively appeared in sub-figures b and c.}
    \vspace{-0.3cm}
    \label{fig:RAID-comparison}
    \vspace{-0.3cm}
\end{figure}

Fig.~\ref{fig:RAID-comparison} compares the availability of three different RAID configurations with equivalent usable (logical) capacity,  
in the presence of human errors.
The results are obtained for a storage by the usable capacity of 21000 disks, for three following configurations: 
a) 7000 $RAID5(3+1)$ arrays, b) 3000 $RAID5(7+1)$ arrays, and c) 21000 $RAID1(1+1)$ arrays.

Comparing the three RAID configurations by assuming no human errors ($hep=0$) shows that $RAID1(1+1)$ results in lower NOMDL compared to 
$RAID5(3+1)$ and $RAID5(7+1)$, while $RAID5(7+1)$ has higher NOMDL compared to $RAID5(3+1)$. 
This observation corroborates our conventional belief that higher redundancy results in higher dependability.  
However, by considering the effect of human errors, we observe $RAID1(1+1)$ configuration shows higher NOMDU compared to both $RAID5$ configurations, while $RAID5(7+1)$ shows the lowest NOMDU. 
This can be described by the higher \emph{Effective Replication Factor\footnote{The ratio of storage physical size to 
the logical (usable) size~\cite{muralidhar2014f4}.}} (ERF) of 
$RAID1(1+1)$ ($ERF=2$) compared to $RAID5(3+1)$ ($ERF=1.33$) and $RAID5(7+1)$ ($ERF=1.14$), which mandates employing higher number of disks for a specific
usable capacity, increasing the chance of disk failure and consequently, human errors.

Another observation is that by increasing $hep$ to 0.01 and beyond, NOMDL caused by DDF in $RAID1(1+1)$ surpasses both $RAID5(7+1)$ and 
$RAID5(3+1)$. It means that in the environments with high probability of human errors, $RAID1$ is not only less available than $RAID5$, but also 
less reliable.  

\subsection{Effect of Automatic Disk Fail-over Policy}
\label{sec:hotSpareReport}

\begin{figure}
    \centering
  	\subfigure[NOMDU]
  	{
        \includegraphics[width=0.35\textwidth]{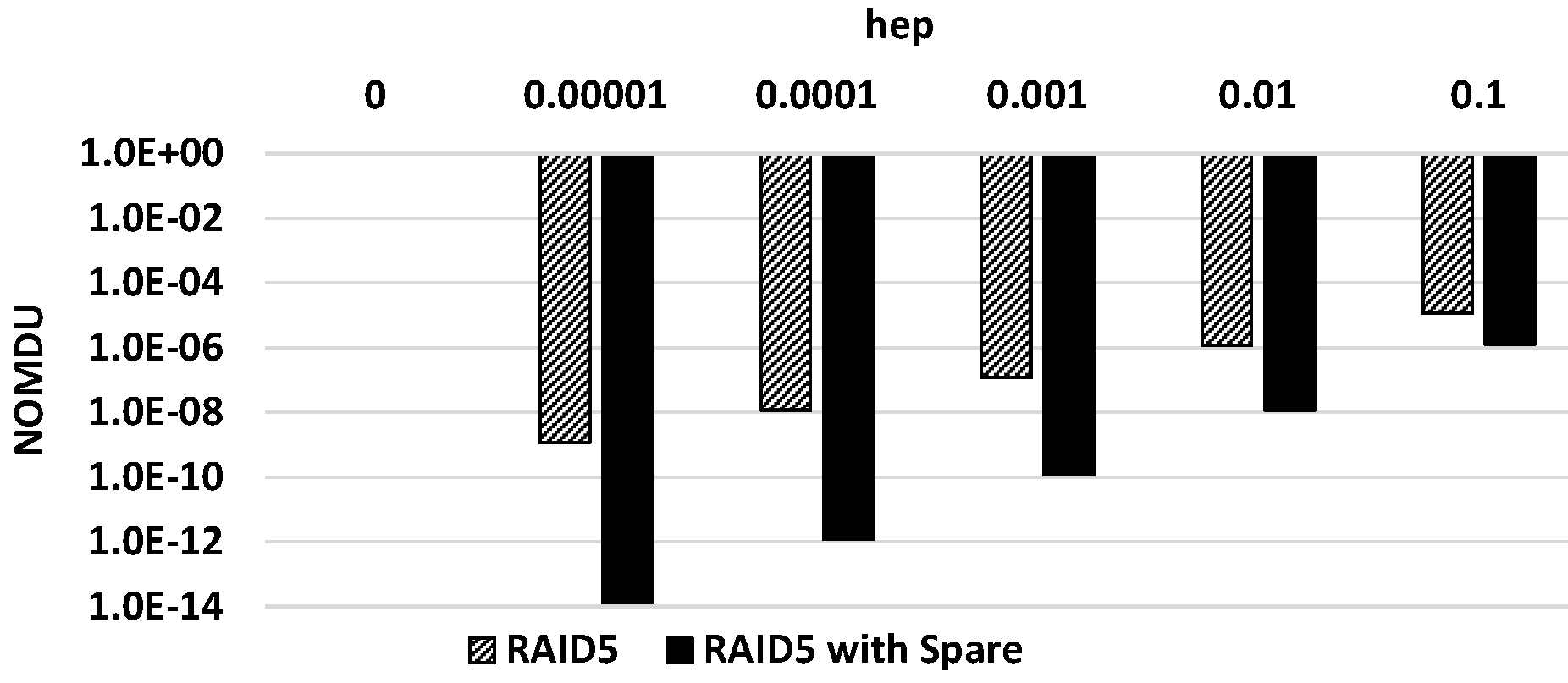}
        \label{fig:spare-disk-nomdu}
        }
    \subfigure[NOMDL-DDF]
    {
    	\includegraphics[width=0.35\textwidth]{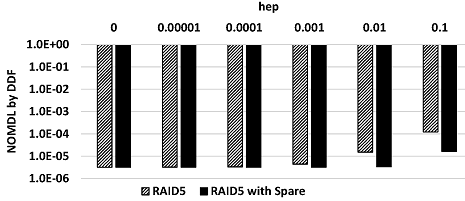}
    	\label{fig:spare-disk-nomdl-ddf}
    	}
    \subfigure[NOMDL-DF+LSE]
    {
    	\includegraphics[width=0.35\textwidth]{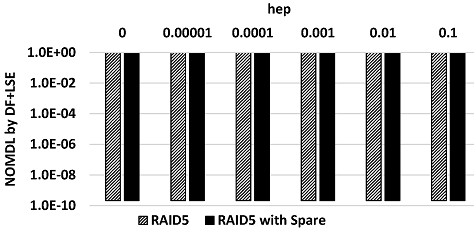}
    	\label{fig:spare-disk-nomdl-lsedf}
    	}
    \caption{NOMDU and NOMDL caused by human errors for conventional $RAID5$ configuration and $RAID5$ with hot spare disk and delayed disk replacement policy. The experiments are conducted for 1000 $RAID5(7+1)$ arrays of Disk A (Table~\ref{tab:weibull-parameters}). We differentiate NOMDL caused by DDF and LSE+DF, respectively appeared in sub-figures b and c.}
    \label{fig:spare-disk}
    \vspace{-0.3cm}
\end{figure}

In this section, we report the effect of the automatic fail-over with hot-spare disk, 
when the service agent follows delayed disk replacement policy, as described in Section~\ref{sec:raid5-spare-lse}.
Fig.~\ref{fig:spare-disk} compares the NOMDU and NOMDL of basic $RAID5$ array and $RAID5$ with hot-spare disk (for 1000 arrays of disk A). 
As the results show, using automatic fail-over policy can significantly moderate the effect of human errors. 
For example, assuming $hep=0.00001$, automatic fail-over decreases NOMDU by five orders of magnitude 
as compared to the conventional RAID.  
Another observation is that automatic fail-over policy can also decrease NOMDL caused by human errors. 
The $hep$ of 0.01 and 0.1 respectively increases NOMDL by 4.7x and 38x compared to the case of no human error, while by using 
automatic fail-over policy, $hep$ of 0.01 and 0.1 increases NOMDL by 1.04x and 5.2x, respectively, as shown in Fig.~\ref{fig:spare-disk-nomdl-ddf}.

\vspace{-0.3cm}
\subsection{Comparison with Previous Models and Field Data}
\label{sec:Comparison with Previous Models and Field Data}
In this section, we compare the results of our proposed model (considering human errors) for $RAID5$ array with the previous $RAID5$ reliability models, including conventional MTTDL model by Gibson~\cite{Gibson-BOOK-1990}, NOMDL by Greenan~\cite{Greenan-HOTSTORAGE-2010}, and DDF by Elerath and Pecht~\cite{Elerath-2009-TC}, where none of them consider the effect of human error and subsequent DU/DL.
Table~\ref{tab:comparison-greenan-elerath} compares previous disk array reliability models with the proposed model for 1000 arrays of $RAID5(7+1)$ and 10 years mission time for Disk A, Disk B, and Disk C. 
In this comparison, we assume a non-survivable storage system (clarified in Section~\ref{sec:raid5-no lse-no spare}) with no spare disk and typical value hep=0.001, while the rest of model parameters is appeared in Table~\ref{tab:weibull-parameters} and Table~\ref{tab:human-error-parameters}. 

As reported in Table~\ref{tab:comparison-greenan-elerath}, only the proposed model considers the effect of human errors and corresponding DU. As an example, the proposed model reports that for $RAID5(7+1)$ arrays of Disk A, 5567 bytes data loss is expected per 1TB of data, in a 10-years mission. It also reports that for $RAID5(7+1)$ arrays of Disk A, 113 bytes are expected to be unavailable per 1TB of data per hour (as NOMDU value is normalized to mission time). NOMDL by Greenan reports that for $RAID5(7+1)$ arrays of Disk A, 4355 bytes data loss is expected per 1TB of data, in a 10-years mission. NOMDL by Greenan is slightly lower, due to the effect of DL caused by human errors considered in our proposed model. 
DDF by Elerath reports that for 1000 $RAID5(7+1)$ arrays of Disk A, 169 DDF incidences happen in a 10-years mission. However, the DDF value has no information about how many of DDFs are caused by DF+DF (that results in the whole array data loss) and how many are caused by DF+LSE (that results in one/multiple stripe data loss). DDF is also a function of examined arrays, 1000 in this case, while NOMDL and NOMDU are normalized to the storage usable capacity and are independent of the number of examined arrays. 
Finally, MTTDL by Gibson reports 8-years mean time to data loss for 1000 $RAID5(7+1)$ arrays of Disk A. This metric has no information about the expected number of failures, the amount of data loss, and the effect of human errors.

\begin{table}[]
\centering
\caption{Comparison of previous disk array reliability models with the proposed model for 1000 arrays of $RAID5(7+1)$ and 10 years mission time. We assume typical value $hep=0.001$ and no spare disk in this comparison, while the rest of model parameters is appeared in Table~\ref{tab:weibull-parameters}, Table~\ref{tab:human-error-parameters}, and Table~\ref{tab:dl-recovery-parameters}. None of previous models consider the effect of human errors on DU/DL.}
\label{tab:comparison-greenan-elerath}
\begin{adjustbox}{width=0.5\textwidth,totalheight=\textheight,keepaspectratio}
\begin{tabular}{|c|c|c|c|c|c|c|}
\hline
\multirow{2}{*}{Disk Array Reliability Model}   & \multicolumn{3}{c|}{DL}                       & \multicolumn{3}{c|}{DU}                                       \\ \cline{2-7} 
                                                & Disk A        & Disk B        & Disk C        & Disk A              & Disk B             & Disk C             \\ \hline
\multirow{2}{*}{NOMDL/NOMDU (Proposed) 10 years} & \multicolumn{3}{c|}{Bytes lost per usable TB} & \multicolumn{3}{c|}{Bytes unavailable per hour per usable TB} \\ \cline{2-7} 
                                                & 5567          & 20871         & 5276          & 113                 & 79                 & 118                \\ \hline
\multirow{2}{*}{NOMDL (Greenan~\cite{Greenan-HOTSTORAGE-2010}) 10 years}        & \multicolumn{3}{c|}{Bytes lost per usable TB} & \multicolumn{3}{c|}{Not considered}                           \\ \cline{2-7} 
                                                & 4355          & 19374         & 4031          & -                   & -                  & -                  \\ \hline
\multirow{2}{*}{DDF (Elerath~\cite{Elerath-2009-TC,elerath-DSN-2007}) 10 years}          & \multicolumn{3}{c|}{Number of DDF incidences} & \multicolumn{3}{c|}{Not considered}                           \\ \cline{2-7} 
                                                & 169           & 35            & 1             & -                   & -                  & -                  \\ \hline
\multirow{2}{*}{MTTDL (Gibson~\cite{Gibson-BOOK-1990}) 10 years}            & \multicolumn{3}{c|}{MTTDL years}              & \multicolumn{3}{c|}{Not considered}                           \\ \cline{2-7} 
                                                & 8            &18            & 17            & -                   & -                  & -                  \\ \hline
\end{tabular}
\end{adjustbox}
\end{table}

\hl{Finally, to further show the shortcoming of previous works in neglecting the effect of human errors, we compare the proposed model results with previous work and field data from enterprise-level storage products of a leading storage system manufacturer and storage service provider (here we call this company by \emph{CorpX}), as shown in Table~\ref{tab:comparison-field-data-table}. Field statistics on the failures of four enterprise-level storage series of this company roughly report that 15\% of all data loss and data unavailability is caused by human errors. 
}

\begin{table*}
\centering
    \caption{\hl{Comparison of the proposed model results with previous work and field data from enterprise-level storage products of a leading storage system manufacturer and storage service provider.}}
\label{tab:comparison-field-data-table}
\begin{adjustbox}{width=0.8\textwidth,totalheight=\textheight,keepaspectratio}
    \begin{tabular}{ | c | c | c |}
    \hline
&	NOMDL	& NOMDU \\ \hline
\textbf{Field Data}	& \textbf{0.00164}	& 15\% of total DU \\ \hline
\textbf{Proposed Model (hep = 0.001)} &	 \textbf{0.00158} &	1.61E-08 \\ \hline
Proposed Model (hep = 0.0001) &	0.00141 &	9.96E-10 \\ \hline
Proposed Model (hep = 0.01) &	0.00316 &	1.58E-07 \\ \hline
Proposed Model (hep = 0.1) &	0.0166 &	1.82E-06 \\ \hline
Greenan~\cite{Greenan-HOTSTORAGE-2010} and Elerath~\cite{Elerath-2009-TC,elerath-DSN-2007} approach considering disk failure and LSE with Weibull distribution (hep = 0.0) &	0.00140 &	0 \\ \hline
Conventional approach considering disk failure with exponential distribution &	0.00145 &	0 \\ \hline
    \end{tabular}
 \end{adjustbox}  
 \vspace{-0.3cm} 

\end{table*}

\hl{As this comparison shows, the DL prediction of Greenan~\cite{Greenan-HOTSTORAGE-2010} and Elerath~\cite{Elerath-2009-TC,elerath-DSN-2007} method is lower than the proposed model, as Greenan and Elerath predict no DL caused by human errors (they just consider DL caused by device failure and LSE). Consequently, total DL reported by the proposed model is 13\% greater than Elerath~\cite{Elerath-2009-TC,elerath-DSN-2007} and Greenan~\cite{Greenan-HOTSTORAGE-2010}. The more significant shortcoming of previous works, however, is ignoring the effect of data unavailability caused by human errors. The \emph{CorpX} field data reports that 15\% of total storage unavailability is caused by human errors, while the previous models do not consider the human error impact by any means. 
Comparing the proposed model results with the field data shows that total DL reported by the proposed model is in the same order with the field data when we choose hep = 0.001 and $\eta_{crash} = 10h$. We are satisfied with this result, as \emph{CorpX} also reports the average human error probability in the same range (0.02\% to 0.1\%). These results are reported for $RAID5(7+1)$ configuration while the field data for other erasure codes are not available. 
The field statistics of Data Loss breakdown, obtained by DeepSpar~\cite{deepspar} (a data recovery firm) from a survey of 50 data recovery firms shows that 12\% of data loss in disk subsystems is caused by human errors~\cite{deepspar}. This statistics is also in the same order with the proposed model results. The proposed model shows 12.8\% of DL is caused by human errors when considering hep=0.001. 
We can conclude that by considering hep = 0.001, the proposed model results are accurate estimate to the field reports. This observation corroborates our previous hep evaluation based on human error statistics from Sharif data-center and related reports on human errors in the field. }

\vspace{-0.3cm}
\subsection{Comparison of Monte Carlo Simulation and Markov Model}
\label{sec:Comparison of Monte Carlo Simulation and Markov Model Results}
\begin{figure}
\begin{centering}
\includegraphics[width=3.3in]{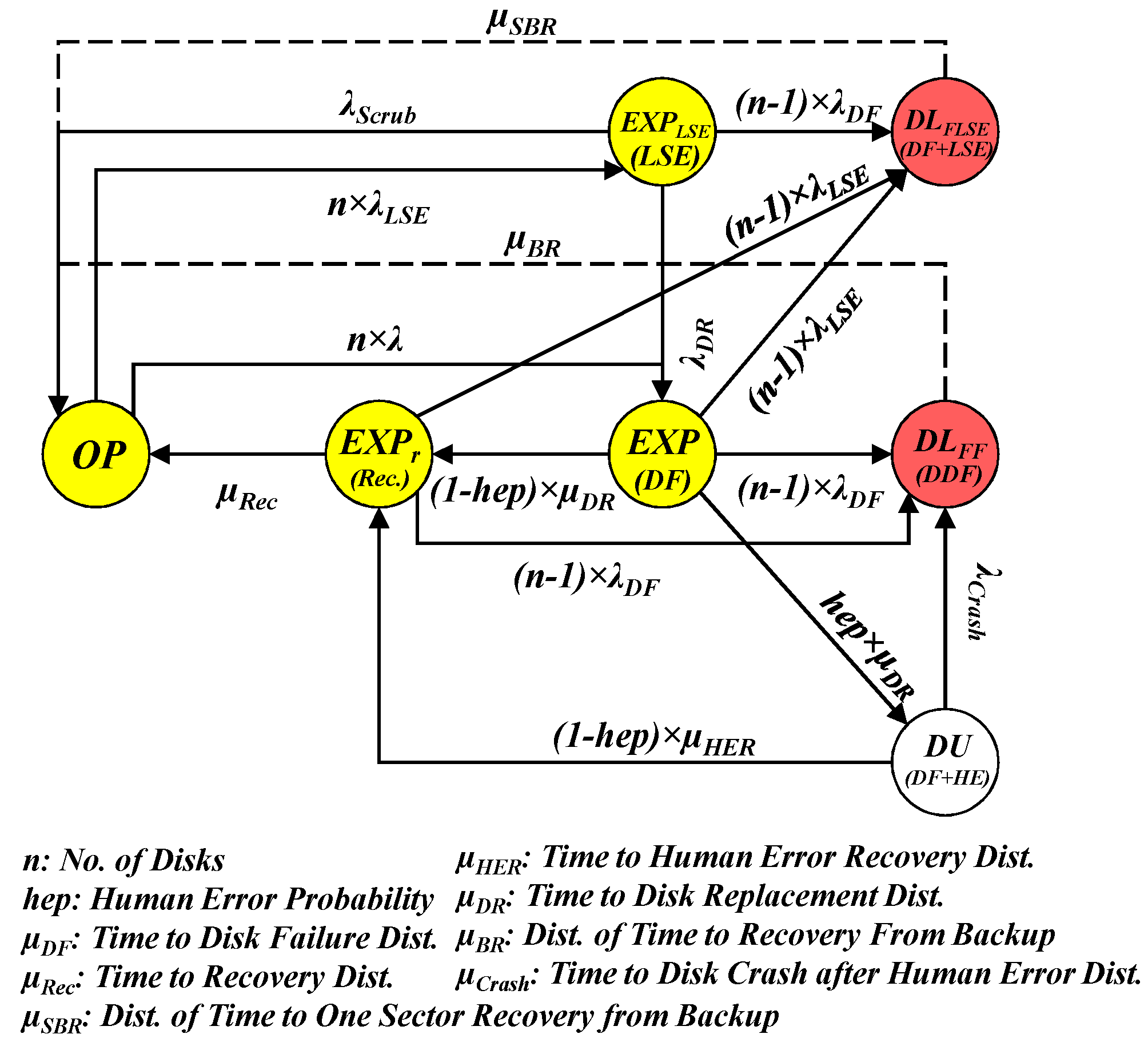}
\caption{Markov Model for $RAID5$ DU/DL, considering LSE.}
\label{fig:raid5markovhrlse}
\par\end{centering}
\vspace{-0.3cm}
\end{figure}

In this section, we compare the results obtained from Markov model with Monte Carlo simulations. In this regard, Markov model of $RAID5$ array (assuming no spare disk and not survivable data, i.e., $DOS(t)=0$) is solved by algebraic approach and then NOMDU and NOMDL are obtained. The Markov model state diagram is same as Monte Carlo simulation state diagram (shown in Fig.~\ref{fig:raid5hrlse}) by considering exponential failure distribution (rather than Weibull distribution used in Monte Carlo simulations), with transition rates appeared in Fig.~\ref{fig:raid5markovhrlse}.
The model parameters are appeared in Table~\ref{tab:weibull-parameters}, Table~\ref{tab:human-error-parameters}, and Table~\ref{tab:dl-recovery-parameters} for Weibull distribution. 

To have a fair comparison between Monte Carlo simulation and Markov models, we justify MTTF/MTTR in exponential distribution to result in the same number of failures as Weibull distribution does in a 10-years mission time. In this regard, both Weibull and exponential distributions should have the same \emph{Cumulative Distribution Function} (CDF) in ten years, as shows in Equation~\ref{equ:cdf}.
\vspace{-0.3cm}

\begin{equation}
\label{equ:cdf}
\begin{split}
F_{exponential}(t) = e^{-MTTF \times t}
\\ F_{Weibull}(t)=e^{-(\frac{t}{\eta})^{\beta}}
\\F_{Weibull}(t)= F_{Exponential}(t) \to MTTF = \frac{(\frac{t}{\eta})^{\beta}}{t}
\end{split}
\end{equation}

\begin{figure}
    \centering
	\subfigure[NOMDU]
	{  
        \includegraphics[width=0.4\textwidth]{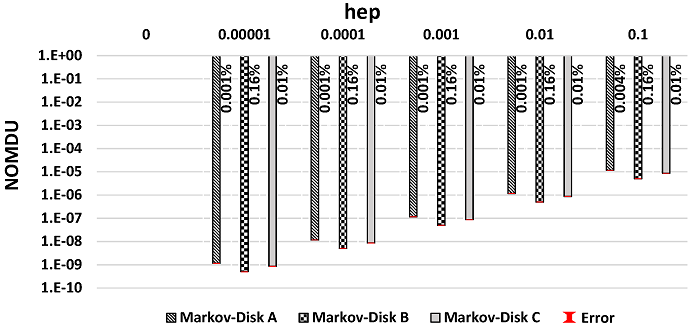}
        \label{fig:nomdu-markov}
        }
    \subfigure[NOMDL-DDF]
    {
    	\includegraphics[width=0.4\textwidth]{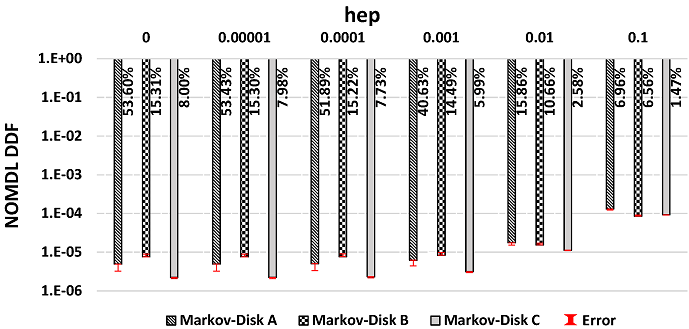}
    	\label{fig:nomdl-ddf-markov}
    	}
  	\subfigure[NOMDL-DF+LSE]
  	{
  		\includegraphics[width=0.4\textwidth]{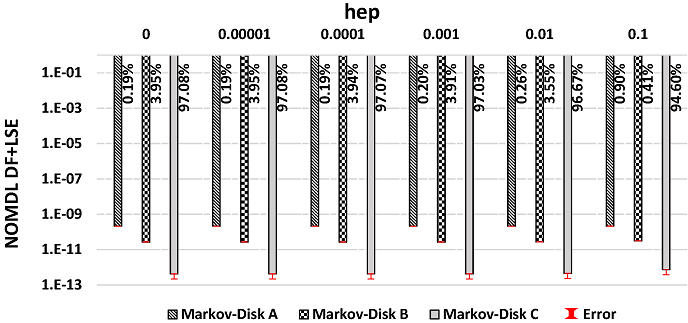}
  		\label{fig:nomdl-dflse-markov}
  		}
     \vspace{-0.3cm}
    \caption{Comparison between Monte Carlo simulation and Markov model results. The NOMDU and NOMDL obtained from Markov model is reported for different \emph{hep} for 1000 $RAID5(7+1)$ arrays of Disk A, Disk B, and Disk C (Table~\ref{tab:weibull-parameters}). The error bar is drawn with respect to Monte Carlo simulation results. The error percentage also appears beside each bar. We differentiate NOMDL caused by DDF and LSE+DF, respectively appeared in sub-figures b and c.}
    \label{fig:markov-mc-comparison}
\end{figure}

Where $t$ is time, $\eta$ is characteristic life, $\beta$ is shape parameter, and $MTTF$ is Mean Time to Failure.
$MTTR$ is obtained by the same equation. Then we set $t$ to 10 years (87600 hours) and calculate $MTTF$ and $MTTR$ of exponential distribution. As such, both Weibull and exponential distributions generate the same number of failure/repair incidences (disk failure, LSE, disk repair, and scrubbing) within 10 years mission time.
Fig.~\ref{fig:markov-mc-comparison} shows Markov model results and the error of Markov model with respect to Monte Carlo simulation results. The error bar (in red color) and error percentage (appeared beside each bar) is also included in this figure. As the figure shows, Markov results have up to 97\% error (in NOMDL DF+LSE for Disk C), while the lowest error is observed in NOMDU (less than 0.1\% for all three disks and 0.05\% on average). However, NOMDL DDF has average error of 37\%, 13\%, and 6\% respectively for disk A, disk B, and disk C (average of 19\% for all three disks). NOMDL DF+LSE has also an average error of 0.3\%, 3\%, and 97\% respectively for disk A, disk B, and disk C (average of 33\% for all three disks). Hence, the highest error of NOMDL DF+LSE belongs to disk C, while the highest error of NOMDL DDF belongs to disk A and the highest error of NOMDU belongs to disk B.

\vspace{-0.3cm}
\subsection{\hl{Model Results For Global Erasure Codes}}
\hl{In this section, we examine the dependability of general erasure codes presented in Section~\ref{sec:general-erasure-codes}.
 In addition to $RAID5$ ($PMDS(m,n,1,0)$) and $RAID6$ ($PMDS(m,n,2,0)$), here we examine $PMDS(m,n,1,1)$, $PMDS(m, n,1,2)$, and $PMDS(m,n,2,2)$, by considering the effect of disk failures, LSEs, and human errors. We choose $PMDS(m,n,1,1)$ and $PMDS(m,n,1,2)$ that have a slightly greater ERF than $RAID5$, but considerably lower ERF than $RAID6$. Both $PMDS(m,n,1,1)$ and $PMDS(m,n,1,2)$ can cope with one device failure and respectively one and two symbol failures (due to respectively having one and two Global parities). $PMDS(m,n,2,2)$ has a ERF greater than both $RAID5$ and $RAID6$, while it can cope with two device failures alongside two symbol failures per code-word.    
}

\begin{figure}
    \centering

        \includegraphics[width=0.5\textwidth]{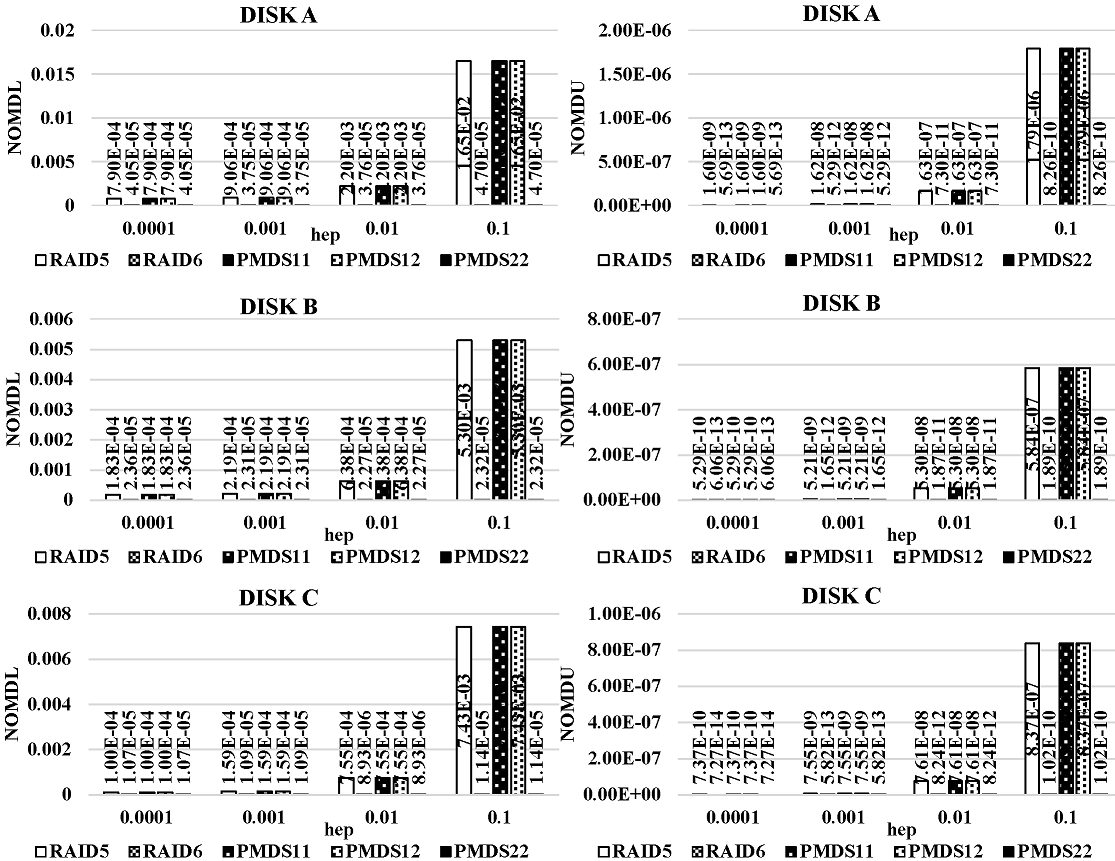}
    \caption{\hl{NOMDU and NOMDL obtained by Monte Carlo simulations for different configurations of PMDS codes.}}
    \label{fig:nomdu-nomdl-pmds}
\end{figure}

\hl{Using the framework described in Section~\ref{sec:monte carlo simulation}, we conduct Monte Carlo simulations and check the failure conditions appeared in Table~\ref{tab:pmds-failure-conditions} to recognize $ADL$, $SDL$, $ADU$, and $SDU$ failure cases and finally calculate NOMDU and NOMDL. }
\hl{In summary, by considering ADL, SDL, ADU, and SDU statistics, we obtain NOMDU and NOMDL as shown in Fig.~\ref{fig:nomdu-nomdl-pmds}.
One important observation in the NOMDU and NOMDL results of different erasure codes is that the codes with the same number of row parities have almost the same NOMDL and NOMDU value. We can justify this observation by the fact that the magnitude of data unavailability and magnitude of data loss caused by device failures is significantly greater than stripe failures. In specific, per ADL event, the magnitude of data loss is 8TB (assuming 1TB disks and array size of 8), versus 128KB per SDL event (hence, the magnitude of ADL is 62,500,000 times greater than SDL). This fact results in the superiority of the effect of ADL and ADU events in the final NOMDU and NOMDL values. For example, NOMDL of $RAID6$ and $PMDS(2,2)$ is very similar ($4.05249887\times 10^{-5}$ and $4.0524983\times 10^{-5}$, respectively), as both arrays perform the same in ADU and ADL, but different in SDU and SDL, due to having the same number of row parities and different number of global parities.  We can also observe that in all erasure codes, human error increases both NOMDL and NOMDU by almost one order of magnitude that corroborates our previous observations on $RAID5$. 
}

\vspace{-0.2cm}
\section{Conclusion and Future Works}
\label{sec:Conclude}
In this paper, we investigated the effect of incorrect disk replacement service on the data unavailability and data loss of 
disk subsystem by using Monte Carlo simulations. 
We also proposed NOMDU, as the duration of data unavailability multiplied to the logical amount of unavailable data, normalized to the mission time and logical capacity of storage system, as a more useful availability metric for storage systems. 
By taking the effect of incorrect disk replacement service into account, 
it is shown that human errors can cause the unavailability of storage array by order of magnitude. 
The human error can also increase the probability of data loss, specially when the human error probability is greater than 0.01. 
It is also shown that in case the human error probability is high (0.01 and beyond), the conventional 
dependability ranking of RAID configurations is contradicted.  
Lastly, the model results show that automatic fail-over can significantly decrease the data unavailability and data loss, caused by human errors, by orders of magnitude. 
Such information can be employed by both designers and system administrators to increase the system dependability.
\vspace{-0.2cm}

\ifCLASSOPTIONcaptionsoff
  \newpage
\fi



%

\bibliographystyle{IEEEtran} 
\bibliography{mss-ras-bib}

\begin{IEEEbiography}[{\includegraphics[width=1in,height=1in,clip]{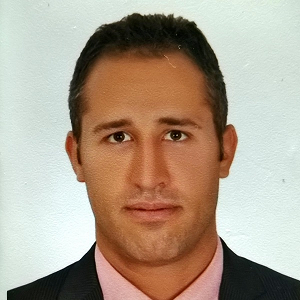}}]%
{Mostafa Kishani}
received the B.S. degree in computer engineering from Ferdowsi University of Mashhad, Mashhad, Iran, in 2008, and M.S. degree in computer 
Engineering from Amirkabir University of Technology (AUT), Tehran, Iran, in 2010. 
He is currently a PhD student of computer engineering in the Sharif University of Technology (SUT), Tehran, Iran, since 2012.
He was a hardware engineer in Iranian Space Research Center (ISRC) from 2010 to 2012.
He was also a member of Institute for Research in Fundamental Sciences (IPM) Memocode team in 2010.
From September 2015 to April 2016 he was a research assistant in Computer Science and Engineering department of the Chinese University of 
Hong Kong (CUHK), Hong Kong.
He was also a research associate in the Hong Kong Polytechnic University (PolyU), Hong Kong, from April 2016 to February 2017.

\end{IEEEbiography}

\begin{IEEEbiography}[{\includegraphics[width=1in,height=1.25in,clip]{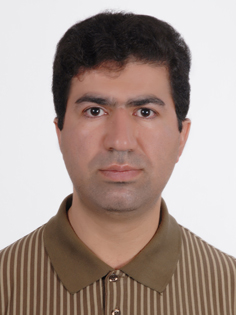}}]%
{Hossein Asadi}
(M'08, SM'14) received the B.Sc. and M.Sc. degrees in computer engineering from the SUT, Tehran, Iran, in 2000 and 2002, respectively, and the Ph.D. degree in electrical and computer engineering from Northeastern University, Boston, MA, USA, in 2007. 

He was with EMC Corporation, Hopkinton, MA, USA, as a Research Scientist and Senior Hardware Engineer, from 2006 to 2009. From 2002 to 2003, he was a member of the Dependable Systems Laboratory, SUT, where he researched hardware verification techniques. From 2001 to 2002, he was a member of the Sharif Rescue Robots Group. He has been with the Department of Computer Engineering, SUT, since 2009, where he is currently a tenured Associate Professor. He is the Founder and Director of the \emph{Data Storage, Networks, and Processing} (DSN) Laboratory, Director of Sharif \emph{High-Performance Computing} (HPC) Center, the Director of Sharif \emph{Information and Coummnications Technology Center} (ICTC), and the President of Sharif ICT Innovation Center. He spent three months in the summer 2015 as a Visiting Professor at the School of Computer and Communication Sciences at the Ecole Poly-technique Federele de Lausanne (EPFL). He is also the co-founder of HPDS corp., designing and fabricating midrange and high-end data storage systems. He has authored and co-authored more than eighty technical papers in reputed journals and conference proceedings. His current research interests include data storage systems and networks, solid-state drives, operating system support for I/O and memory management, and reconfigurable and dependable computing.

Dr. Asadi was a recipient of the Technical Award for the Best Robot Design from the International RoboCup Rescue Competition, organized by AAAI and RoboCup, a recipient of Best Paper Award at the 15th CSI Internation Symposium on \emph{Computer Architecture \& Digital Systems} (CADS), the Distinguished Lecturer Award from SUT in 2010, the Distinguished Researcher Award and the Distinguished Research Institute Award from SUT in 2016, and the Distinguished Technology Award from SUT in 2017. He is also recipient of Extraordinary Ability in Science visa from US Citizenship and Immigration Services in 2008. He has also served as the publication chair of several national and international conferences including CNDS2013, AISP2013, and CSSE2013 during the past four years. Most recently, he has served as a Guest Editor of IEEE Transactions on Computers, an Associate Editor of Microelectronics Reliability, a Program Co-Chair of CADS2015, and the Program Chair of CSI National Computer Conference (CSICC2017). 
\end{IEEEbiography}

\newpage

\appendices
\label{sec:appendices}

\section{\hl{Dependability Analysis of General Erasure Codes}}
\label{sec:appendix-dependability-analysis}
\hl{
\subsection{ADL Condition}
ADL happens in a very simple condition, when the number of failed devices ($DF$) surpasses $r$ (the number of redundant devices).
}
\begin{equation}
\begin{centering}
r < DF
\end{centering}
\end{equation}

\vspace{-0.5cm}
\hl{
\subsection{SDL Condition}
SDL happens when ADL condition is not satisfied, but there exist at least one stripe where the number of LSEs surpasses the maximum correctable LSEs. Stripe $v$ has the following number of LSEs:
}
\begin{equation}
\begin{centering}
\sum_{i=1}^{n} NUM_{LSE}(i,v)
\end{centering}
\end{equation}
\hl{The maximum correctable LSEs per stripe is the aggregation of LSEs correctable by global parities and LSEs correctable by row parities. The number of LSEs correctable by global parity is equal to $s$ (number of global parities). However, the number of LSEs correctable by row parity depends on the number of failed devices ($DF$) and the distribution of LSEs in the stripe. Using $PMDS(m,n,r,s)$, in each stripe we can behave $h$ number of operational devices as failed device and correct all their LSEs using row parities, where:} 
\begin{equation}
\begin{centering}
h=r-DF
\end{centering}
\end{equation}
\hl{$h$, is the number of operational devices that are behaved as failed device and all of their LSEs (regardless of the number of LSEs in that device) are corrected using row parities. To attain the maximum possible correction capability, we select $h$ devices that have the maximum number of LSEs. Hence, the maximum correctable LSEs using row parities is as follows:}
\begin{equation}
\begin{centering}
\sum_{i=1}^{r-DF}NUM_{LSE}(MAX(i,v),v)
\end{centering}
\end{equation}
\hl{Finally, SDL happens when the following condition is satisfied:}
\begin{equation}
\begin{centering}
\resizebox{1\hsize}{!}{$(DF \leq r) \wedge (\exists v \in V [s + \sum_{i=1}^{r-DF} NUM_{LSE}(MAX(i,v),v) < \sum_{i=1}^{n} NUM_{LSE}(i,v)])$}
\end{centering}
\end{equation}

\hl{
\subsection{ADU Condition}
ADU happens when ADL condition is not satisfied, but the aggregation of failed devices ($DF$) and unavailable devices by human error ($HE$) surpasses $r$:
}
\begin{equation}
\begin{centering}
(DF \leq r) \wedge (r<DF+HE) 
\end{centering}
\end{equation}

\hl{
\subsection{SDU Condition}
SDU happens when ADU and ADL conditions are not satisfied and at least one stripe exists where the number of LSEs does not surpass the maximum correctable LSEs, but its data is unavailable due to human error. For satisfying SDU condition, at least one human error is happened and ADU and ADL conditions are unsatisfied:}
\begin{equation}
\begin{centering}
(0 < HE) \wedge (DF + HE \leq r) 
\end{centering}
\end{equation}
\hl{Moreover, the stripe $v$ has no lost sectors under the following condition (as discussed in the case of SDL):}
\begin{equation}
\begin{centering}
 \sum_{i=1}^{n} NUM_{LSE}(i,v) - \sum_{i=1}^{r-DF} NUM_{LSE}(MAX(i,v),v) \leq s
\end{centering}
\end{equation}
\hl{Finally, stripe $v$ has unavailable sectors under the condition that the number of LSEs in the available devices does not surpass the maximum LSEs obtainable with the available devices. The number of LSEs in the available devices is as follows:}
\begin{equation}
\begin{centering}
\sum_{i=1}^{n} NUM_{LSE}(i,v) \times OP(i)
\end{centering}
\end{equation}
\hl{Maximum LSEs obtainable with available devices is the aggregation of LSEs obtainable with global parities and LSEs obtainable with row parities. The number of LSEs obtainable by global parity is equal to $s$ (the number of global parities). However, the number of LSEs obtainable by row parities is a function of the number of failed devices ($DF$), number of unavailable devices due to human error ($HE$), and the distribution of LSEs in the stripe. Using $PMDS(m,n,r,s)$, in each stripe we can behave $h$ number of operational devices as unavailable device and obtain all their LSEs using row parities (regardless of the number of LSEs in that device), where:}
\begin{equation}
\begin{centering}
h = r - DF - HE
\end{centering}
\end{equation}
\hl{To obtain the maximum possible LSEs, we select $h$ operational devices that have the maximum number of LSEs. Hence, the maximum obtainable LSEs using row parities is as follows:}
\begin{equation}
\begin{centering}
\sum_{i=1}^{r-DF-HE} NUM_{LSE}(MAXOP(i,v),v)
\end{centering}
\end{equation}
\hl{And the maximum obtainable LSEs in stripe $v$ is the aggregation of $s$ and above value. Hence, stripe $v$ has unavailable sectors under the following condition:}
\begin{equation}
\begin{centering}
\resizebox{1\hsize}{!}{$s + \sum_{i=1}^{r-DF-HE}NUM_{LSE}(MAXOP(i,v),v) < \sum_{i=1}^{n}NUM_{LSE}(i,v) \times OP(i)$}
\end{centering}
\end{equation}
\hl{All in all, SDU happens when the following condition is satisfied:}
\begin{equation}
\begin{split}
\resizebox{1\hsize}{!}{$(0<HE) \wedge (DF+HE \leq r) \wedge (\exists v \in V [(\sum_{i=1}^{n}NUM_{LSE}(i,v) \leq s + \sum_{i=1}^{r-DF}NUM_{LSE}(MAX(i,v),v))$} 
\\ \resizebox{0.85\hsize}{!}{$\wedge (s + \sum_{i=1}^{r-DF-HE}NUM_{LSE}(MAXOP(i,v),v) < \sum_{i=1}^{n}NUM_{LSE}(i,v) \times OP(i))]) $}
\end{split}
\end{equation}


\vspace{-0.5cm}
\section{\hl{Cumulative Number of ADL and SDL Incidences}}
\label{sec:appendix-cumulative-adl-sdl}
\begin{figure}
    \centering
	\subfigure[Normal Disk Size]
	{  
        \includegraphics[width=0.45\textwidth]{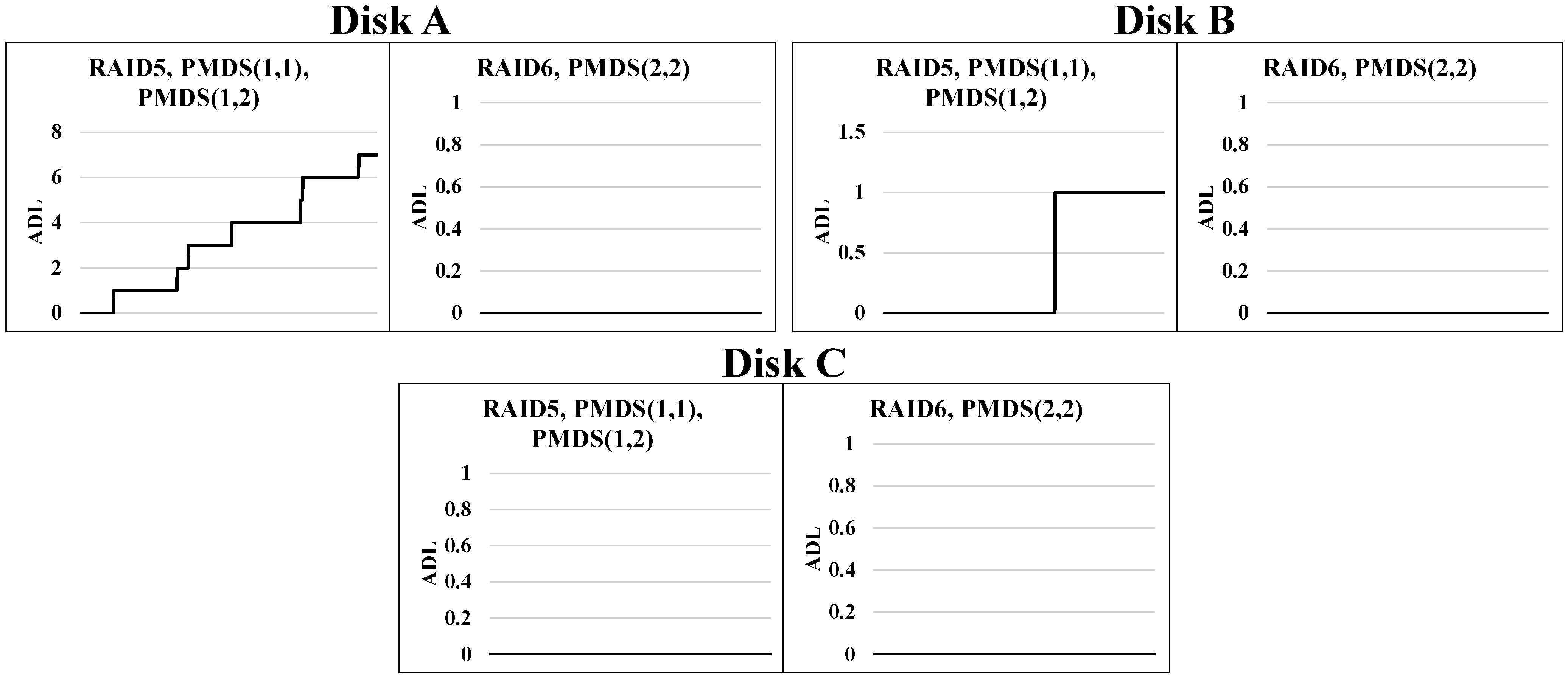}
        \label{fig:adl-normal}
        }
    \subfigure[Small Disk Size]
    {
    	\includegraphics[width=0.45\textwidth]{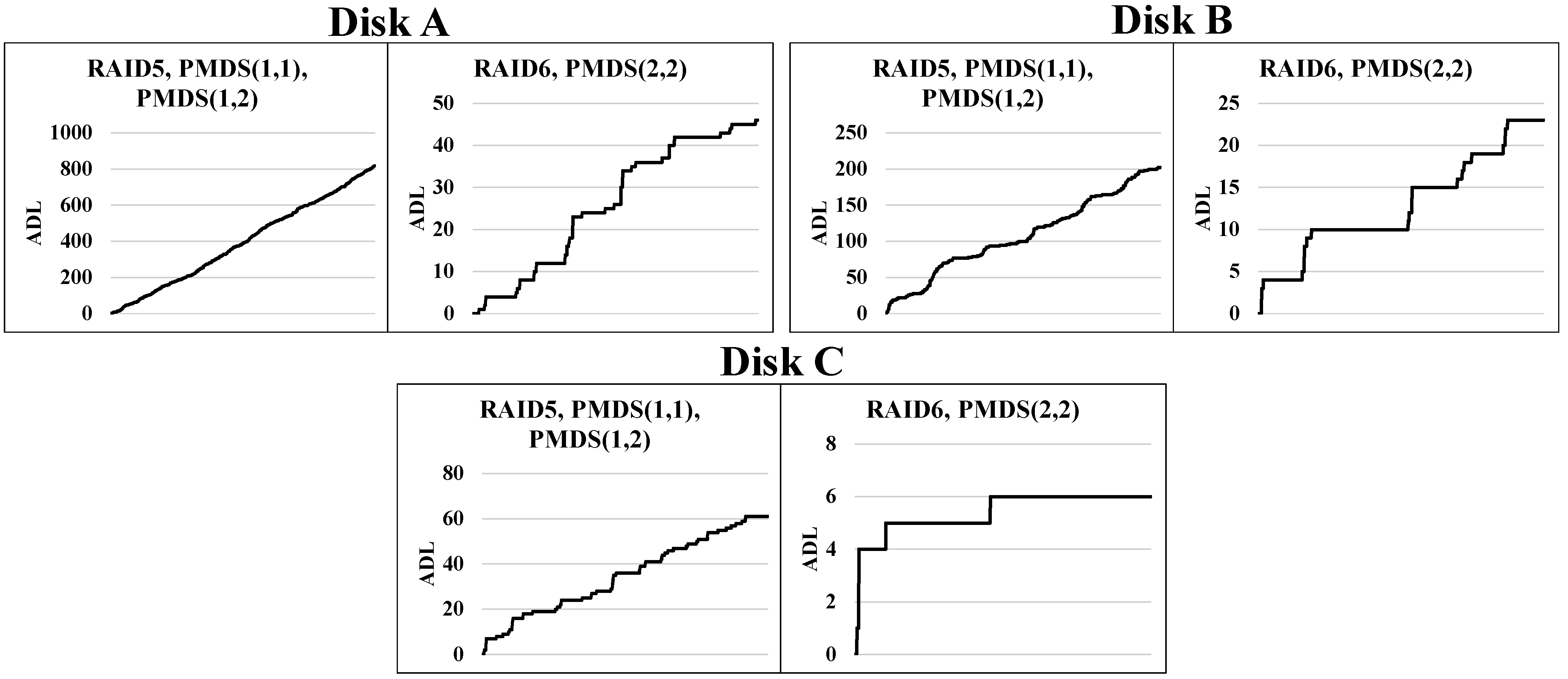}
    	\label{fig:adl-small}
    	}
  \subfigure[Ultra-Small Disk Size]
  {
  	\includegraphics[width=0.45\textwidth]{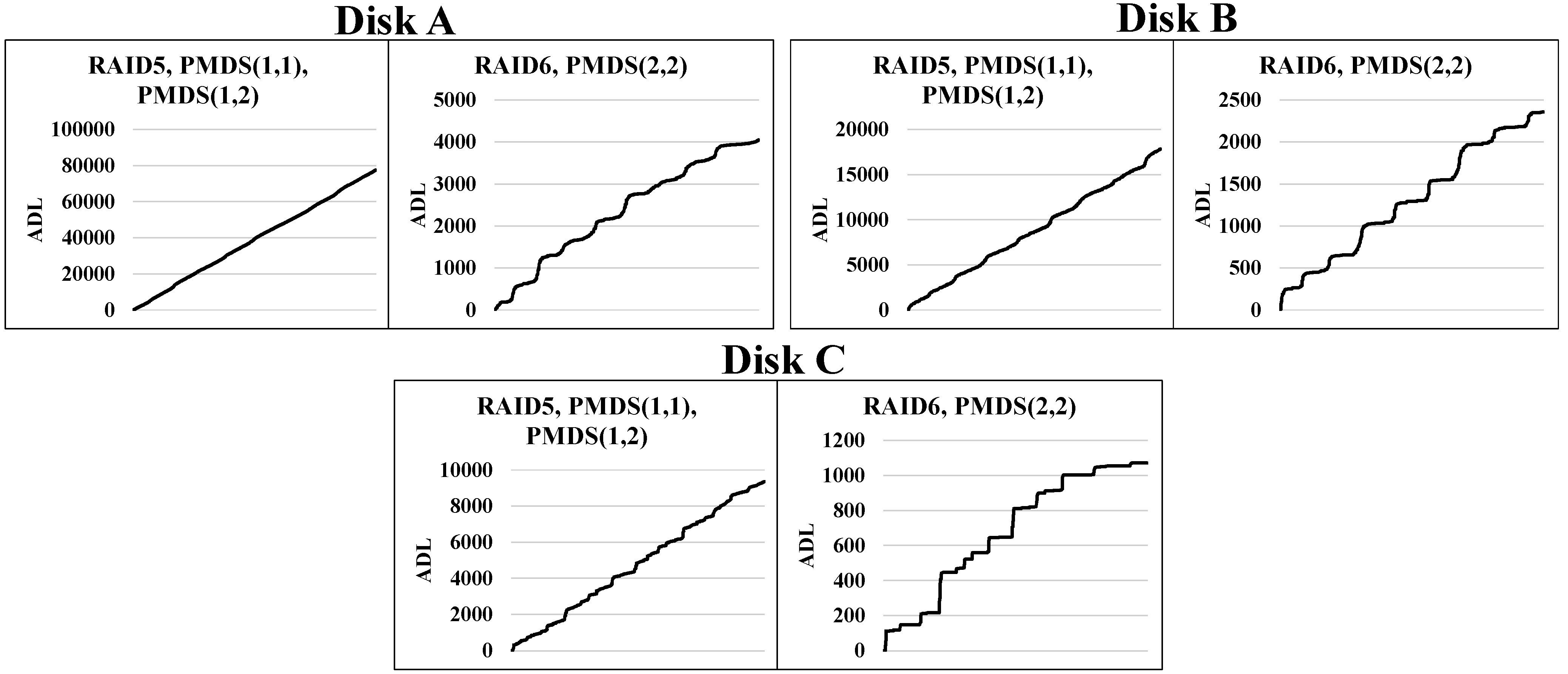}
  	\label{fig:adl-ultra-small}
  	}
    \caption{\hl{Accumulative ADL obtained by Monte Carlo simulations for different configurations of PMDS codes.}}
    \label{fig:adl}
\end{figure}

\begin{figure}
    \centering
	\subfigure[Normal Disk Size]
	{  
        \includegraphics[width=0.5\textwidth]{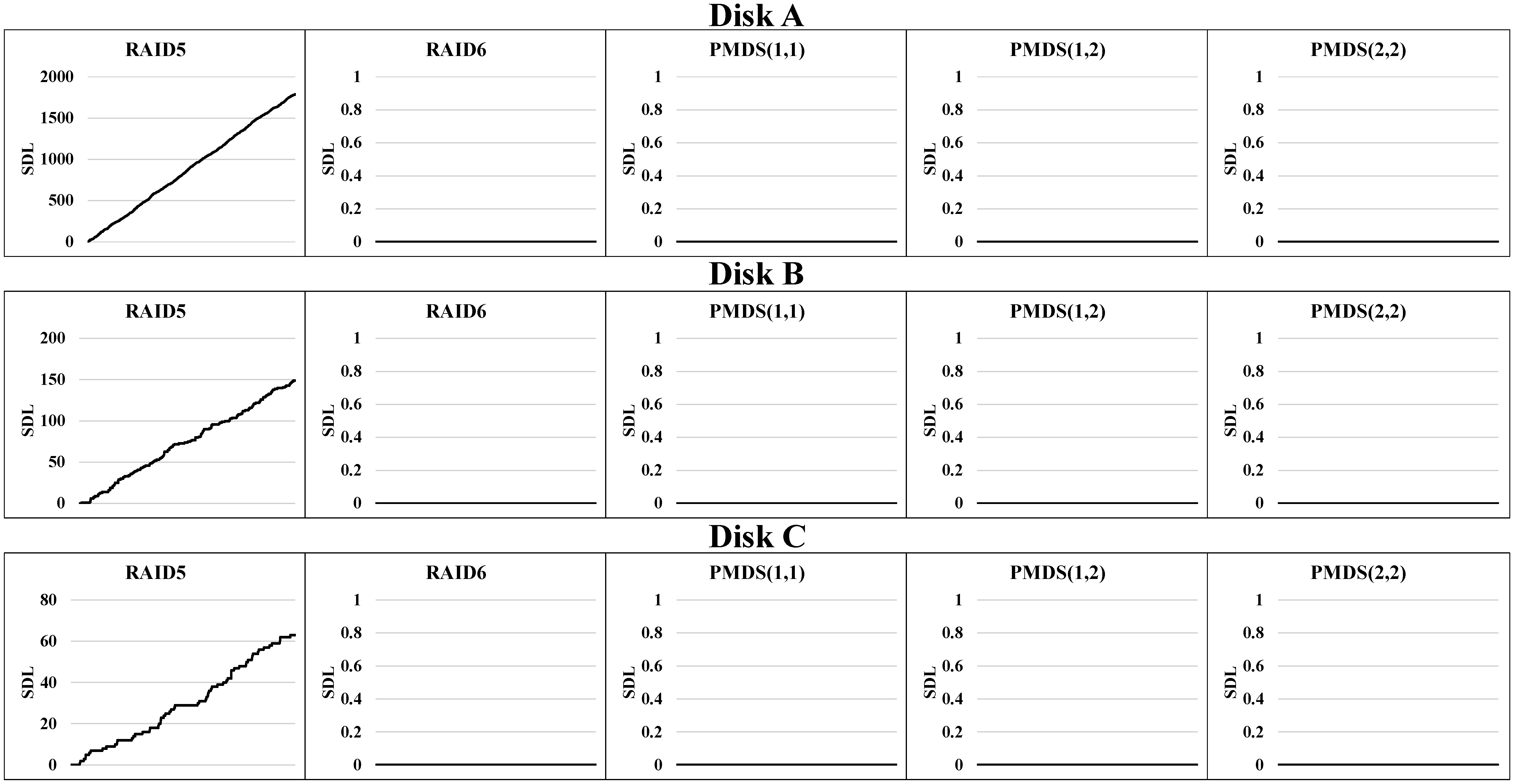}
        \label{fig:sdl-normal}
        }
    \subfigure[Small Disk Size]
    {
    	\includegraphics[width=0.5\textwidth]{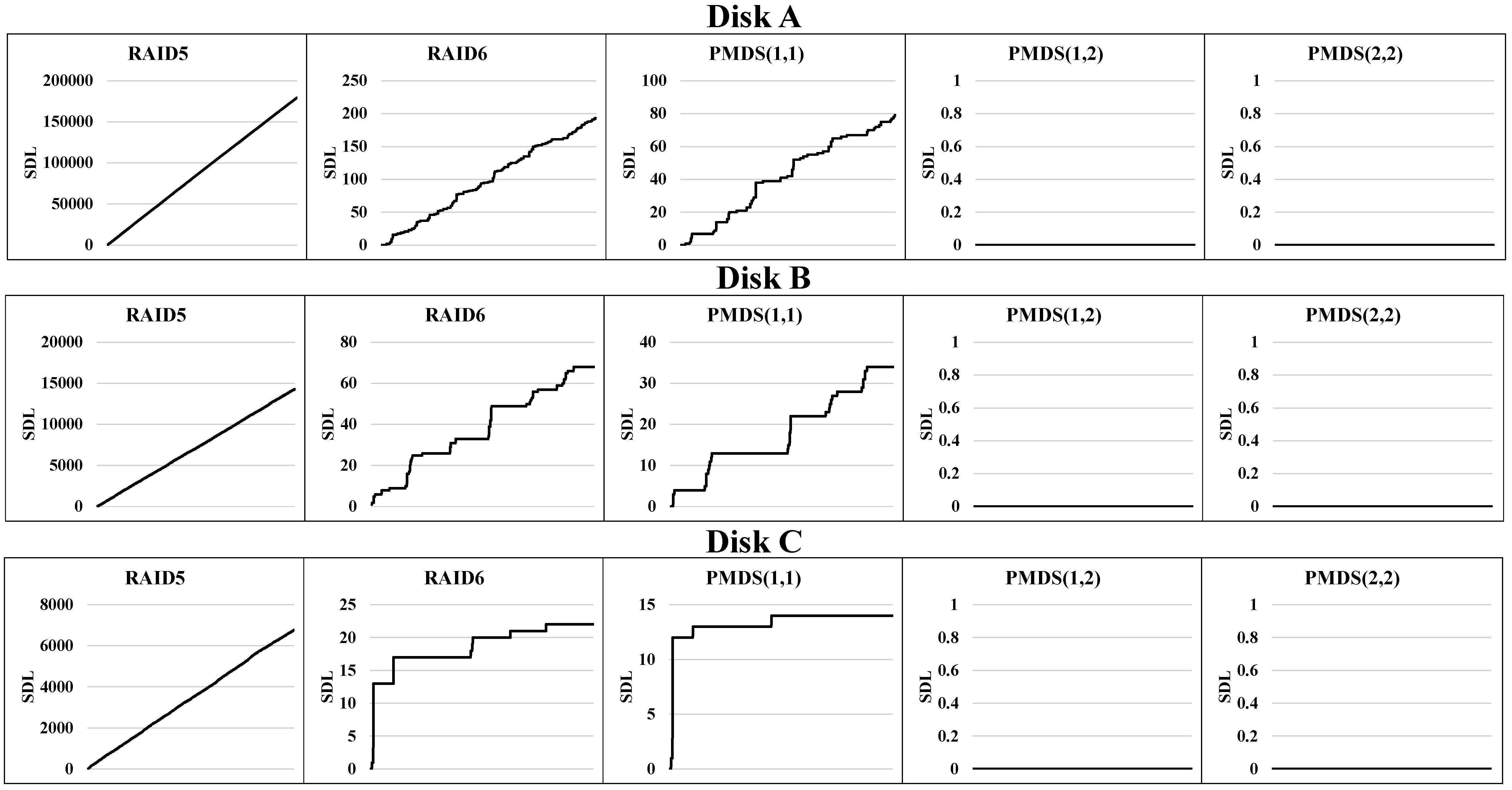}
    	\label{fig:sdl-small}
    	}
  	\subfigure[Ultra-Small Disk Size]
  	{
  		\includegraphics[width=0.5\textwidth]{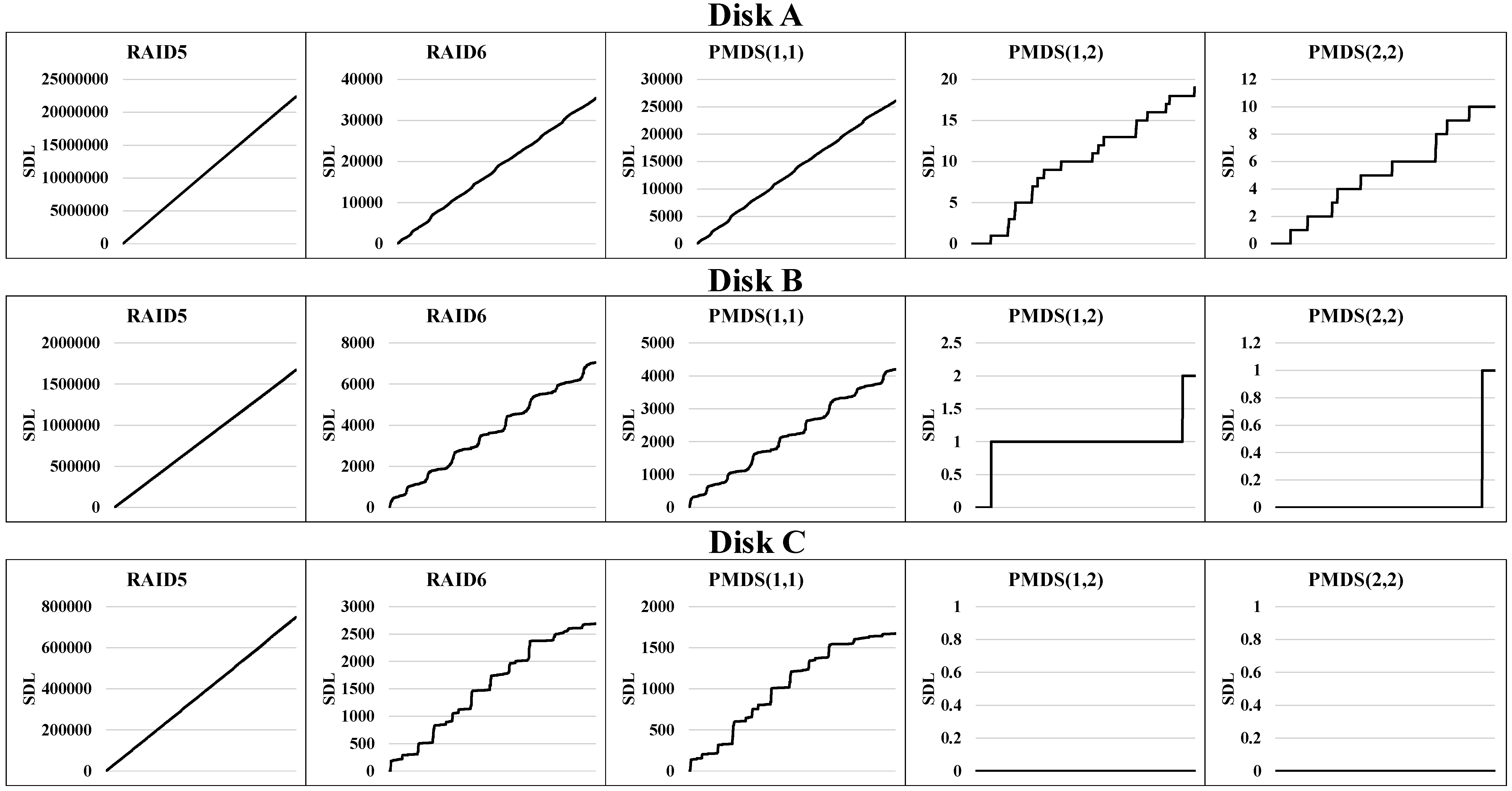}
  		\label{fig:sdl-ultra-small}
  		}
    \caption{\hl{Accumulative SDL obtained by Monte Carlo simulations for different configurations of PMDS codes.}}
    \label{fig:sdl}
\end{figure}

\hl{In Fig.~\ref{fig:adl} and Fig.~\ref{fig:sdl}, we respectively draw the cumulative number of ADL and SDL incidences for $RAID5(7+1)$, $RAID6(7+2)$, $PMDS(m,8,1,1)$,  $PMDS(m,8,1,2)$, and $PMDS(m,9,2,2)$, respectively denoted as $RAID5$, $RAID6$, $PMDS(1,1)$, $PMDS(1,2)$, and $PMDS(2,2)$ in the charts. To have a fair comparison, the erasure codes are considered to have almost equal usable capacity of seven drives (note PMDS codes have a usable capacity slightly lower than 7, due to the overhead of Global Parities). The simulation parameters are appeared in Table~\ref{tab:weibull-parameters}, Table~\ref{tab:human-error-parameters}, and Table~\ref{tab:dl-recovery-parameters}. As the number of ADL incidences depends on the number of row parity devices, the erasure codes with the same number of row parities result in the same number of ADL in each fault injection experiment. Hence, we concatenate the ADL of $RAID5$, $PMDS(1,1)$, and $PMDS(1,2)$, and also concatenate ADL of $RAID6$ and $PMDS(2,2)$ in Fig.~\ref{fig:adl}.}

\hl{The first set of results is obtained for 10,000 disk arrays working for 10 years (87600 hours) considering the real capacity of each disk (Disk A: 1TB, Disk B: 1TB, Disk C: 288GB), shown in Fig.~\ref{fig:adl-normal} and Fig.~\ref{fig:sdl-normal}.
As we see in the first set of results, the failure cases such as multiple LSEs in the same stripe and triple device failure are so rare. Hence, in practice we see no difference between the results of $RAID6$, $PMDS(1,1)$, $PMDS(1,2)$, and $PMDS(2,2)$. To increase the chance of such failure cases, we decrease the disk sizes by the factor of 64X (we call it \emph{small disk size}). Decreasing the disk size also decreases the simulation time, which makes simulating larger number of disk arrays practical. }

\hl{Fig.~\ref{fig:adl-small} and Fig.~\ref{fig:sdl-small} respectively show the ADL and SDL for 1,000,000 disk arrays with small size. 
In the results obtained by small disk size, we can apparently observe the superiority of $PMDS(1,2)$ and $PMDS(2,2)$ in preventing SDL events (zero number of SDL in our experiments), due to employing two global parities that cope with two sector failures per stripe. The results also show that $PMDS(1,1)$ outperforms $RAID6$ in handling sector failures. For example in the case of disk A, $PMDS(1,1)$ encounters 79 SDL events versus 193 SDL events observed in $RAID6$ array. In the case of array data loss, however, the number of ADL events is a function of employed row parities (employed redundant disks). Hence, we can see that $RAID6$ and $PMDS(2,2)$ outperform the rest of codes by almost one order of magnitude, due to employing two redundant devices rather than one redundant device in $RAID5$, $PMDS(1,1)$, and $PMDS(1,2)$. For example in the case of disk A, $RAID6$ and $PMDS(2,2)$ encounter 46 ADL events versus 818 ADL events observed in the case of $RAID5$, $PMDS(1,1)$, and $PMDS(1,2)$. }

\hl{
Finally, the results of Fig.~\ref{fig:adl-ultra-small} and Fig.~\ref{fig:sdl-ultra-small} are obtained by decreasing the disk sizes by the factor of 16384 (we call it \emph{ultra-small disk size}) for 100,000,000 disk arrays. 
In the results obtained by ultra-small disks, we can further observe the superiority of $PMDS(2,2)$ over $PMDS(1,2)$ in handling sector failures. For example in the case of disk A, we observed 10 SDL events in $PMDS(2,2)$ versus 19 SDL events in $PMDS(1,2)$, as shown in Fig.~\ref{fig:sdl-ultra-small}.
}

%








\end{document}